\documentclass[final,twocolumn]{IEEEtran}

\usepackage{amsmath,epsfig,amssymb,algorithm,amsthm,cite,url}

\usepackage{algpseudocode}
\usepackage{hyperref}

\newtheorem{theorem}{Theorem}
\newtheorem{lemma}[theorem]{Lemma}
\newtheorem{assumption}{Assumption~A-\kern-0pt}

\newtheorem{corollary}[theorem]{Corollary}
\newtheorem{remark}{Remark}
\newtheorem*{remark*}{Remark}
\DeclareMathOperator{\tr}{tr}
\newcommand{\bh}{{\bf h}}
\newcommand{\bH}{{\bf H}}
\newcommand{\bHh}{{\bf H}^{\mbox{\tiny H}}}
\newcommand{\wbHh}{\widehat{\bf H}^{\mbox{\tiny H}}}
\newcommand{\wbH}{\widehat{\bf H}}
\newcommand{\bhh}{{\bf h}_k^{\mbox{\tiny H}}}
\newcommand{\wbh}{\widehat{\bf h}}

\newcommand{\bx}{{\bf x}}
\newcommand{\bz}{{\bf z}}
\newcommand{\bQ}{{\bf Q}}

\newcommand{\var}{{\rm var}}

\newcommand{\C}{{\mathbb C}}
\newcommand{\bG}{{\bf G}}
\newcommand{\bPhi}{\boldsymbol{\Phi}}
\newcommand{\Tau}{\boldsymbol{\mathcal{T}}}
\usepackage{pgfplots}

\newcommand{\maximize}[1]{{\underset{{#1}}{\mathrm{maximize}}}}

\newcommand*\rfrac[2]{{}^{#1}\!\!/_{#2}}
\title{Linear Precoding Based on Polynomial Expansion: Reducing Complexity in Massive MIMO} 

\author{Axel M\"uller,~\IEEEmembership{Student Member,~IEEE,}
	Abla Kammoun,~\IEEEmembership{Member,~IEEE,}
        Emil~Bj\"ornson,~\IEEEmembership{Member,~IEEE,}
        and~M\'erouane~Debbah,~\IEEEmembership{Senior~Member,~IEEE}
\thanks{A.~M\"uller, E.~Bj\"ornson, and M.~Debbah are with the Alcatel-Lucent Chair on Flexible Radio, SUPELEC, Gif-sur-Yvette, France (e-mail: \{axel.mueller, emil.bjornson, merouane.debbah\}@supelec.fr, abla.kammoun@gmail.com).
A.~Kammoun is with KAUST, Thuwal, Makkah Province, Saudi Arabia.
E.~Bj\"ornson is also with the Department of Signal Processing, ACCESS Linnaeus Centre, KTH Royal Institute of Technology, Stockholm, Sweden.}
\thanks{E.~Bj\"ornson is funded by the International Postdoc Grant 2012-228 from The Swedish Research Council. This research has been supported by the ERC Starting Grant 305123 MORE (Advanced Mathematical Tools for Complex Network Engineering). Part of this paper was presented at the $8$th IEEE Sensor Array and Multichannel Signal Processing Workshop, 2014.}
}

\begin{document}

\maketitle

\begin{abstract}
Massive multiple-input multiple-output (MIMO) techniques have the potential to bring tremendous improvements in spectral efficiency to future communication systems. Counter-intuitively, the practical issues of having uncertain channel knowledge, high propagation losses, and implementing optimal non-linear precoding are solved more-or-less automatically by enlarging system dimensions. However, the computational precoding complexity grows with the system dimensions. For example, the close-to-optimal and relatively ``antenna-efficient'' regularized zero-forcing (RZF) precoding is very complicated to implement in practice, since it requires fast inversions of large matrices in every coherence period. Motivated by the high performance of RZF, we propose to replace the matrix inversion and multiplication by a truncated polynomial expansion (TPE), thereby obtaining the new TPE precoding scheme which is more suitable for real-time hardware implementation and significantly reduces the delay to the first transmitted symbol. The degree of the matrix polynomial can be adapted to the available hardware resources and enables smooth transition between simple maximum ratio transmission and more advanced RZF.

By deriving new random matrix results, we obtain a deterministic expression for the asymptotic signal-to-interference-and-noise ratio (SINR) achieved by TPE precoding in massive MIMO systems. Furthermore, we provide a closed-form expression for the polynomial coefficients that maximizes this SINR. To maintain a fixed per-user rate loss as compared to RZF, the polynomial degree does not need to scale with the system, but it should be increased with the quality of the channel knowledge and the signal-to-noise ratio.
\end{abstract}


\section{Introduction}

The current wireless networks must be greatly densified to meet the exponential growth in data traffic and number of user terminals (UTs) \cite{CISCO2013}. The conventional densification approach is to decrease the inter-site distance by adding new base stations (BSs) \cite{Hoydis2011c}. However, the cells are subject to more interference from neighboring cells as distances shrink, which requires substantial coordination between neighboring BSs or fractional frequency reuse patterns. Furthermore, serving high-mobility UTs by small cells is very cumbersome due to the large overhead caused by rapidly recurring handover.

Massive MIMO techniques, also known as large-scale multi-user MIMO techniques, have been shown to be viable alternatives and complements to small cells \cite{Marzetta2010a,Rusek2013a,Hoydis2013a,Hosseini2013a,Bjornson2013e}. By deploying large-scale arrays with very many antennas at current macro BSs, an exceptional array gain and spatial precoding resolution can be obtained. This is exploited to achieve higher UT rates and serve more UTs simultaneously. In this paper, we consider the single-cell downlink case where one BS with $M$ antennas serves $K$ single-antenna UTs. As a rule-of-thumb, hundreds of BS antennas may be deployed in the near future to serve several tens of UTs in parallel. If the UTs are selected spatially to have a very small number of common scatterers, the user channels naturally decorrelate as $M$ grows large \cite{Gao2011a,Hoydis2012a} and space-division multiple access (SDMA) techniques become robust to channel uncertainty \cite{Marzetta2010a}.

One might imagine that by taking $M$ and $K$ large, it becomes terribly difficult to optimize the system throughput. The beauty of massive MIMO is that this is not the case: simple linear precoding is asymptotically optimal in the regime $M \gg K \gg 0$ \cite{Marzetta2010a} and random matrix theory can provide simple deterministic approximations of the stochastic achievable rates \cite{HAC06,Nguyen2008a,WAG10,Muharar2011a,Hoydis2013a,COUbook}. These so-called \emph{deterministic equivalents} are tight as $M$ grows large due to channel hardening, but are usually also very accurate at small values of $M$ and $K$.

Although linear precoding is computationally more efficient than its non-linear alternatives, the complexity of most linear precoding schemes is still intractable in the large-($M,K$) regime since the number of arithmetic operations is proportional to $K^2 M$. For example, both the optimal precoding parametrization in \cite{Bjornson2012c} and the near-optimal \emph{regularized zero-forcing (RZF)} precoding \cite{PEE05} require an inversion of the Gram matrix of the joint channel of all users---this matrix operation has a complexity proportional to $K^2M$. A notable exception is the matched filter, also known as \emph{maximum ratio transmission (MRT)} \cite{Lo1999a}, whose complexity only scales as $MK$. Unfortunately, this precoding scheme requires roughly an order of magnitude more BS antennas to perform as well as RZF \cite{Hoydis2013a}. Since it makes little sense to deploy an advanced massive MIMO system and then cripple the system throughput by using interference-ignoring MRT, treating the precoding complexity problem is the main focus of this paper.

Similar complexity issues appear in multi-user detection, where the minimum mean squared error (MMSE) detector involves matrix inversions \cite{Moshavi1996a}. This uplink problem has received considerable attention in the last two decades; see \cite{Moshavi1996a,Honig2001a,Sessler2005a,hoydis} and references therein. In particular, different reduced-rank filtering approaches have been proposed, often based on the concept of \emph{truncated polynomial expansion (TPE)}. Simply speaking, the idea is to approximate the matrix inverse by a matrix polynomial with $J$ terms, where $J$ needs not to scale with the system dimensions to maintain a certain approximation accuracy \cite{Honig2001a}. TPE-based detectors admit simple and efficient multistage/pipelined hardware implementation \cite{Moshavi1996a}, which stands in contrast to the complicated implementation of matrix inversion. A key requirement to achieve good detection performance at small $J$ is to find good coefficients for the polynomial. This has been a major research challenge because the optimal coefficients are expensive to compute \cite{Moshavi1996a}. Alternatives based on appropriate scaling \cite{Sessler2005a} and asymptotic analysis \cite{hoydis} have been proposed. A similar TPE-based approach was used in \cite{Shariati2013a} for the purpose of low-complexity channel estimation in massive MIMO systems.

In this paper, which follows our work in \cite{Mueller2014b}, we propose a new family of low-complexity linear precoding schemes for the single-cell multi-user downlink.
We exploit TPE to enable a balancing of precoding complexity and system throughput. A main analytic contribution is the derivation of deterministic equivalents for the achievable user rates for any order $J$ of TPE precoding. These expressions are tight when $M$ and $K$ grow large with a fixed ratio, but also provide close approximations at small parameter values. The deterministic equivalents allow for optimization of the polynomial coefficients; we derive the coefficients that maximize the throughput. We note that this approach for precoding design is very new. The only other work is \cite{zarei} by Zarei~\emph{et al.}, of which we just became aware at the time this paper was first submitted. Unlike our work, the precoding in \cite{zarei} is conceived to minimize the sum-MSE of all users. Although our approach builds upon the same TPE concept as \cite{zarei}, the design method proposed herein is more efficient since it considers the optimization of the throughput. This metric is usually more pertinent than the sum-MSE. Additionally, our work is more comprehensive in that we consider a channel model which takes into account the transmit correlation at the base station.

Our novel TPE precoding scheme enables a smooth transition in performance between MRT ($J=1$) and RZF ($J=\min(M,K)$), where the majority of the gap is bridged for small values of $J$. We show that $J$ is independent of the system dimensions $M$ and $K$, but must increase with the signal-to-noise ratio (SNR) and channel state information (CSI) quality to maintain a fixed per-user rate gap to RZF. We stress that the polynomial structure provides a green radio approach to precoding, since it enables energy-efficient multistage hardware implementation as compared to the complicated/inefficient signal processing required to compute conventional RZF. Also, the delay to the first transmitted symbol is significantly reduced, which is of great interest in systems with very short coherence periods. Furthermore, the hardware complexity can be easily tailored to the deployment scenario or even changed dynamically by increasing and reducing $J$ in high and low SNR situations, respectively.

\subsection{Notation}

Boldface (lower case) is used for column vectors, ${\bf x}$, and (upper case) for matrices, ${\bf X}$. Let ${\bf X}^{\mbox{\tiny T}}$, ${\bf X}^{\mbox{\tiny H}}$, and ${\bf X}^{*}$  denote the transpose, conjugate transpose, and conjugate of ${\bf X}$, respectively, while $\tr ({\bf X})$ is the matrix trace function.
The Frobenius norm is denoted $\|\cdot\|$ and the spectral norm is denoted $\|\cdot\|_2$. A circularly symmetric complex Gaussian random vector ${\bf x}$ is denoted ${\bf x} \sim \mathcal{CN}(\bar{{\bf x}},{\bf Q})$, where $\bar{{\bf x}}$ is the mean and ${\bf Q}$ is the covariance matrix.
The set of all complex numbers is denoted by $\C$, with $\C^{N\times 1}$ and $\C^{N\times M}$ being the generalizations to vectors and matrices, respectively. The $M\times M$ identity matrix  is written as ${\bf I}_M$ and the zero vector of length $M$ is denoted ${\bf 0}_{M\times 1}$. For an infinitely differentiable monovariate function $f(t)$, the $\ell$th derivative at $t=t_0$ (i.e., $ \rfrac{d^\ell}{d t^\ell}f(t)|_{t=t_0} $) is denoted by $f^{(\ell)}(t_0)$ and more concisely $f^{(\ell)}$, when $t=0$. An analog definition  is considered in the bivariate case; in particular $f^{(l,m)}(t_0,u_0)$ refers to the $\ell$th and $m$th derivative with respect to $t$ and $u$ at $t_0$ and $u_0$, respectively (i.e., $ \rfrac{\partial^\ell}{\partial t^\ell}\, \rfrac{\partial^m}{\partial u^m}f(t,u)|_{t=t_0, u=u_0} $). If $t_0=u_0=0$ we abbreviate again as $f^{(l,m)}= f^{(l,m)}(0,0)$.
Furthermore, we use the big-$O$ and small-$o$ notation in their usual sense; that is, $\alpha_M = \mathcal{O}(\beta_M)$ serves a flexible abbreviation for $|\alpha_M|\leq C \beta_M$, where $C$ is a generic constant, and $\alpha_M = o(\beta_M)$ is shorthand for $\alpha_M = \varepsilon_M\beta_M$ with $\varepsilon_M\to 0$, as $M$ goes to infinity.

\section{System Model}
This section defines the single-cell system with flat-fading channels, linear precoding, and channel estimation errors.

\subsection{Transmission Model}

We consider a single-cell downlink system in which a BS, equipped with $M$ antennas, serves $K$ single-antenna UTs.
The received complex baseband signal $y_k \in \C$ at the $k$th UT is given by
\begin{equation}
y_k=\bhh \bx + n_k, \quad k=1,\ldots,K
\label{eq:system_model}
\end{equation}
where $\bx \in \C^{M\times 1}$ is the transmit signal and $\bh_k \in \C^{M\times 1}$ represents the random channel vector between the BS and the $k$th UT.
The additive circularly-symmetric complex Gaussian noise at the $k$th UT is denoted by $n_k \sim \mathcal{CN}(0,\sigma^2)$ for $k=1,\ldots,K$, where $\sigma^2$ is the receiver noise variance.

The small-scale channel fading is modeled as follows.

\begin{assumption}
	\label{ass:channel}
	The channel vector ${\bf h}_k$ is modeled as
\begin{equation} \label{eq:channel-model}
	\bh_k =\bPhi^{\frac{1}{2}}\bz_k
\end{equation}
	where the channel covariance matrix $\bPhi \in \mathbb{C}^{M \times M}$ has bounded spectral norm $\|\bPhi\|_2$, as $M \rightarrow \infty$, and $\bz_k \sim \mathcal{CN}({\bf 0}_{M\times 1},{\bf I}_M)$.
The channel vector has a fixed realization for a coherence period and then takes a new independent realization.
This model is known as \emph{Rayleigh block-fading}.
\end{assumption}

Note that we assume that the UTs reside in a rich scattering environment described by the covariance matrix $\bPhi$. This matrix can either be a scaled identity matrix as in \cite{Marzetta2010a} or describe array-specific properties (e.g., non-isotropic radiation patterns) and general propagation properties of the coverage area (e.g., for practical sectorized sites). We  consider a common covariance matrix $\bPhi$ here, as the main focus in this paper it the precoding scheme. This simplification has been done in many recent publications. Adhikary et~al.\ \cite{adhikary2013joint} have proposed to always only serve groups of UTs that share approximately equal covariance matrices, hence providing further motivation behind Assumption \mbox{A-\ref{ass:channel}}.

The application of TPE precoding to multi-cell systems can be found in our paper \cite{Kammoun2014b}. However, the models used in this paper and in \cite{Kammoun2014b} are incompatible and differ most prominently in the assumption whether the total transmit power increases with the number of users as in \cite{Kammoun2014b} or is fixed as in this paper; see \eqref{eq:power-constraint}. 
This seemingly negligible change has a big impact on the analysis and applicability of the models, as this assumption means that the noise term in \cite{Kammoun2014b} become asymptotically zero, while in the current work the noise term is non-negligible. The channel estimation model in \cite{Kammoun2014b} and in this paper are also different and the calculations follow very different approaches, due to the inclusion of power control later on.
Another big extension in the current work is the complete complexity analysis of the TPE approach in comparison to the classical RZF approach. Only this analysis gives TPE precoding its motivation and pertinence.
Finally, we want to point out that the optimization in \cite{Kammoun2014b} is with respect to a max min SNR problem and the solution is not given as a closed form, while here we maximize the throughput and find a closed form solution.
Before utilizing our work, one needs to decide which model gives the most accurate asymptotic behavior for the specific type of system considered.

\begin{assumption}
The BS employs Gaussian codebooks and linear precoding, where ${\bf g}_k \in \C^{M\times 1}$ denotes the precoding vector and $s_k \sim \mathcal{CN}(0,1)$ is the data symbol of the $k$th UT.
\label{ass:linear-precoding}
\end{assumption}

Based on this assumption, the transmit signal in \eqref{eq:system_model} is
\begin{equation}
{\bx} = \sum_{n=1}^{K} {\bf g}_n s_n = \bG {\bf s}.
\label{eq:precoder_fun}
\end{equation}
The matrix notation is obtained by letting $\bG = [ {\bf g}_1 \, \ldots \, {\bf g}_K ] \in \C^{M \times K}$ be the precoding matrix and ${\bf s} = [s_1 \, \ldots \, s_{K}]^{\mbox{\tiny T}} \sim \mathcal{CN}({\bf 0}_{K\times 1},{\bf I}_K) $ be the vector containing all UT data symbols.

Consequently, the received signal \eqref{eq:system_model} can be expressed as
\begin{equation}
y_k= \bhh{\bf g}_k s_k + \sum_{n=1, n\neq k}^K \bhh {\bf g}_n s_n + n_k.
\end{equation}
Let $\bG_k \in \C^{M \times (K-1)}$ be the matrix $\bG$ with column ${\bf g}_k$ removed.
Then the SINR at the $k$th UT becomes
\begin{equation}
{\rm SINR}_k=\frac{\bhh{\bf g}_k{\bf g}_k^{\mbox{\tiny H}}\bh_k}{\bhh\bG_k\bG_k^{\mbox{\tiny H}}\bh_k+\sigma^2}.
\label{eq:sinr}
\end{equation}
By assuming that each UT has perfect instantaneous CSI, the achievable data rates at the UTs are
\begin{equation*}
r_k = \log_2( 1+ {\rm SINR}_k ), \quad k=1,\ldots,K.
\end{equation*}

\subsection{Model of Imperfect Channel Information at Transmitter}

Since we typically have $M \geq K$ in practice, we assume that we either have a time-division duplex (TDD) protocol where the BS acquires channel knowledge from uplink pilot signaling \cite{Hoydis2013a} or a frequency-division duplex (FDD) protocol where temporal correlation is exploited as in \cite{Choi2014a}. In both cases, the transmitter generally has imperfect knowledge of the instantaneous channel realizations and we model this by the generic Gauss-Markov formulation; see
\cite{Wang2006a,WAG10,Nosrat2011a}:
\begin{assumption} The transmitter has an imperfect channel estimate
\begin{equation} \label{eq:uncertainty-model}
     \wbh_k=\bPhi^{\frac{1}{2}}\left(\sqrt{1-\tau^2}\bz_k+\tau {\bf v}_k\right)= \sqrt{1-\tau^2}{\bf h}_k+\tau{\bf n}_k
\end{equation}
for each UT, $k=1,\ldots, K$,
where ${\bf h}_k$ is the true channel, ${\bf v}_k \sim \mathcal{CN}({\bf 0}_{M\times 1},{\bf I}_M)$, and ${\bf n}_k = \bPhi^{\frac{1}{2}} {\bf v}_k \sim \mathcal{CN}({\bf 0}_{M\times 1},\bPhi)$ models the independent error. The scalar parameter $\tau \in [0,1]$ indicates the quality of the instantaneous CSI, where $\tau=0$ corresponds to perfect instantaneous CSI and $\tau=1$ corresponds to having only statistical channel knowledge.
\label{ass:channel_imperfect}
\end{assumption}
The parameter $\tau$ depends on factors such as time/power spent on pilot-based channel estimation and user mobility. Note that we assume for simplicity that the BS has the same quality of channel knowledge for all UTs.

Based on the model in \eqref{eq:uncertainty-model}, the matrix
\begin{equation} \label{eq:uncertainty-model-joint}
\widehat{\bf H}=\left[\widehat{\bf h}_1 \, \ldots \, \widehat{\bf h}_K\right] \in \mathbb{C}^{M \times K}
\end{equation}
denotes the joint imperfect knowledge of all user channels.

\section{Linear Precoding}

Many heuristic linear precoding schemes have been proposed in the literature, mainly because finding the optimal precoding (in terms of weighted sum rate or other criteria) is very computational demanding and thus unsuitable for fading systems \cite{Bjornson2013d}. Among the heuristic schemes we distinguish \emph{RZF precoding} \cite{PEE05}, which is also known as transmit Wiener filter \cite{Joham2005a}, signal-to-leakage-and-noise ratio maximizing beamforming \cite{Sadek2007a}, generalized eigenvalue-based beamformer \cite{Stridh2006a}, and virtual SINR maximizing beamforming \cite{Bjornson2010c}. The reason that RZF precoding has been proposed by different authors (under different names) is, most likely, that it provides close-to-optimal performance in many scenarios. It also outperforms classical MRT and zero-forcing beamforming (ZFBF) by combining the respective benefits of these schemes \cite{Bjornson2013d}. Therefore, RZF is deemed the natural starting point for this paper.

Next, we provide a brief review of RZF and prior performance results in massive MIMO systems. These results serve as a starting point for Section~\ref{subsection:polynomial-expansion}, where we propose an alternative precoding scheme with a computational/hardware complexity more suited for large systems.

\subsection{Review on RZF Precoding in Massive MIMO Systems}

Suppose we have a total transmit power constraint
\begin{equation} \label{eq:power-constraint}
 \tr \left( \bG\bG^{\mbox{\tiny H}} \right) = P.
\end{equation}
We stress that the total power $P$ is fixed, while we let the number of antennas, $M$, and number of UTs, $K$, grow large.

Similar to \cite{WAG10}, we define the RZF precoding matrix as
\begin{align}
	\bG_{\rm RZF}&= \frac{\beta}{\sqrt{K}} \widehat{\bH}\left(\frac{1}{K}{\wbHh}\widehat{\bH}+\xi {\bf I}_K\right)^{-1}{\bf P}^{\frac{1}{2}}\nonumber\\
				   &=\beta\left(\frac{1}{K}\widehat{\bH}\wbHh +\xi {\bf I}_M \right)^{-1}\frac{\widehat{\bH}}{\sqrt{K}}{\bf P}^{\frac{1}{2}}\label{eq:RZF}
\end{align}
where 
the power normalization parameter $\beta$ is set such that $\bG_{\rm RZF}$ satisfies the power constraint in \eqref{eq:power-constraint} and ${\bf P}$ is a fixed diagonal matrix whose diagonal elements are power allocation weights for each user. We assume that ${\bf P}$ satisfies:
\begin{assumption}
\label{ass:power}
	The diagonal values $p_k,\ k=1,\ldots,K$ in ${\bf P}={\rm diag}(p_1,\ldots,p_K)$ are positive and of order $\mathcal{O}(\frac{1}{K})$.
\end{assumption}
The scalar regularization coefficient $\xi$ can be selected in different ways, depending on the noise variance, channel uncertainty at the transmitter, and system dimensions \cite{PEE05,WAG10}.
In \cite{WAG10}, the performance of each UT under RZF precoding is studied in the large-$(M,K)$ regime. This means that $M$ and $K$ tend to infinity at the same speed, which can be formalized as follows.
\begin{assumption}
	\label{ass:regime}
In the large-$(M,K)$ regime, $M$ and $K$ tend to infinity such that
	$$
	0< \liminf \frac{K}{M} \leq \limsup \frac{K}{M} < +\infty.
	$$	
\end{assumption}

The user performance is characterized by ${\rm SINR}_k$ in \eqref{eq:sinr}. Although the SINR is a random quantity that depends on the instantaneous values of the random users channels in ${\bf H}$ and the instantaneous estimate $\widehat{\bf H}$, it can be approximated using deterministic quantities in the large-$(M,K)$ regime \cite{HAC06,Nguyen2008a,WAG10,Muharar2011a}. These are quantities that only depend on the statistics of the channels and are often referred to as \emph{deterministic equivalents}, since they are almost surely ($\mathrm{a.s.}$) tight in the asymptotic limit. This channel hardening property is essentially due to the law of large numbers. Deterministic equivalents were first proposed by Hachem~\emph{et~al.} in \cite{HAC06}, who have also shown their ability to capture important system performance indicators. When the deterministic equivalents are applied at finite $M$ and $K$, they are referred to as \emph{large-scale approximations}.

In the sequel, by deterministic equivalent of a sequence of random variables $X_n$ we mean a deterministic sequence $\overline{X}_n$ which approximates $X_n$ such that
\begin{equation}
X_n-\overline{X}_n\xrightarrow[n\to+\infty]{\mathrm{a.s.}}0.
\end{equation}

As an example, we recall the following result from \cite{HAC06}, which provides some widely known results on deterministic equivalents. Note that we have chosen to work with a slightly different definition of the deterministic equivalents than in \cite{HAC06}, since this better fits the analysis of our proposed precoding scheme.

\begin{theorem}[Adapted from \cite{HAC06}]
	\label{th:deterministic}
	Consider the resolvent matrix $\bQ(t)=\left(\frac{t}{K}\bH\bHh+ {\bf I}_M\right)^{-1}$ where the columns of $\bH$ are distributed according to Assumption \mbox{A-\ref{ass:channel}}. Then, the equation
		$$
	\delta(t)=\frac{1}{K}\tr \left( \boldsymbol{\Phi}\left({\bf I}_M+\frac{t\boldsymbol{\Phi}}{1+t\delta(t)}\right)^{-1} \right)
		$$
admits a unique solution $\delta(t)>0$ for every $t>0$.\\
Let ${\bf T}(t)=\left({\bf I}_M+\frac{t\boldsymbol{\Phi}}{1+t\delta(t)}\right)^{-1}$ and let ${\bf U}$ be any matrix with bounded spectral norm.
Under Assumption \mbox{A-\ref{ass:regime}} and for $t>0$, we have
\begin{equation} \label{eq:determinstic-equivalent}
		\frac{1}{K}\tr \left(  {\bf U}\bQ(t)  \right)- \frac{1}{K}\tr  \left( {\bf U}{\bf T}(t) \right)  \xrightarrow[M,K\to+\infty]{\mathrm{a.s.}} 0.
\end{equation}
\end{theorem}

The statement in \eqref{eq:determinstic-equivalent} shows that $\frac{1}{K}\tr({\bf U}{\bf T}(t))$ is a deterministic equivalent to the random quantity $\frac{1}{K}\tr({\bf U}\bQ(t))$.

In this paper, the deterministic equivalents are essential to determine the limit to which the SINRs tend in the large-$(M,K)$ regime.
For RZF precoding, as in \eqref{eq:RZF}, this limit is given by the following theorem.

\begin{theorem}[Adapted from Corollary 1 in \cite{WAG10}] \label{theorem:SINR-limits}
	Let $\rho=\frac{P}{\sigma^2}$ and consider the notation ${\bf T}={\bf T}(\frac{1}{\xi})$ and $\delta=\delta(\frac{1}{\xi})$. Define the deterministic scalar quantities
	$$
	\gamma=\frac{1}{K} \tr \left( {\bf T}\boldsymbol{\Phi}{\bf T}\boldsymbol{\Phi} \right)
	$$
	and
\begin{equation} \label{eq:RZF-theta} 
	\theta =\frac{(1-\tau^2)\frac{p_k}{\tr({\bf P})/K}\delta^2\left((\delta+\xi)^2-\gamma\right)}{\gamma\left(\xi^2-\tau^2(\xi^2-(\xi+\delta)^2)\right)+\frac{1}{K}\tr \left( \boldsymbol{\Phi}{\bf T}^2 \right) \frac{(\xi+\delta)^2}{\rho}}.
\end{equation}
Then, the SINRs with RZF precoding  satisfies
	$$
	{\rm SINR}_k -\theta \xrightarrow[M,K\to+\infty]{\mathrm{a.s.}} 0, \quad k=1,\ldots,K.
	$$
\end{theorem}

Note that all UTs obtain the same asymptotic value of the SINR, since the UTs have homogeneous channel statistics.
Theorem \ref{theorem:SINR-limits} holds for any regularization coefficient $\xi$, but the parameter can also be selected to maximize the limiting value $\theta$ of the SINRs. This is achieved by the following theorem.

\begin{theorem}[Adapted from Proposition 2 in \cite{WAG10}] \label{theorem:optimal-regularization}
	Under the assumption of a uniform power allocation, $p_k=\frac{P}{K}$, the large-scale approximated SINR in \eqref{eq:RZF-theta} under RZF precoding is maximized by the regularization parameter $\xi^\star$, given as the positive solution to the fixed-point equation
	$$
	\xi^\star=\frac{1}{\rho }\frac{1+\nu(\xi^\star)+\tau^2\rho \frac{\gamma}{\frac{1}{K}\tr \left( {\bf T}\boldsymbol{\Phi}^2 \right) }}{(1-\tau^2)(1+\nu(\xi^\star))+\frac{1}{(\xi^\star)^2}\tau^2\nu(\xi^\star)(\xi+\delta)^2}
	$$
	where $\nu(\xi)$ is given by
	$$
	\nu(\xi)=\frac{\xi\frac{1}{K}\tr \left( \boldsymbol{\Phi}{\bf T}^3 \right) }{\gamma \frac{1}{K}\tr \left( \boldsymbol{\Phi}{\bf T}^2 \right) }\left(\frac{\gamma}{\frac{1}{K}\tr \left( \boldsymbol{\Phi}{\bf T}^2 \right) }-\frac{\frac{1}{K} \tr \left( \boldsymbol{\Phi}^2{\bf T}^3 \right) }{\frac{1}{K}\tr \left( \boldsymbol{\Phi}{\bf T}^3 \right)}\right).
	$$
\end{theorem}

The RZF precoding matrix in \eqref{eq:RZF} is a function of the instantaneous CSI at the transmitter. Although the SINRs converges to the deterministic equivalents given in Theorem \ref{theorem:SINR-limits}, in the large-$(M,K)$ regime, the precoding matrix remains a random quantity that is typically recalculated on a millisecond basis (i.e., at the same pace as the channel knowledge is updated). This is a major practical issue, because the matrix inversion operation in RZF precoding is very computationally demanding in large systems \cite{Shepard2012a}; the number of operations scale as $\mathcal{O}(K^2 M)$ and the known inversion algorithms are complicated to implement in hardware (see Section~\ref{section:complexity-analysis} for details). The matrix inversion is the key to interference suppression in RZF precoding, thus there is need to develop less complicated precoding schemes that still can suppress interference efficiently.

\subsection{Truncated Polynomial Expansion Precoding}
\label{subsection:polynomial-expansion}
Motivated by the inherent complexity issues of RZF precoding, we now develop a new linear precoding class that much easier to implement in large systems.
The precoding is based on rewriting the matrix inversion by a polynomial expansion, which is then truncated. The following lemma provides a major motivation behind the use of polynomial expansions.

\begin{lemma} \label{lemma:inversion-expansion}
For any positive definite Hermitian matrix ${\bf X}$,
\begin{equation}
{\bf X}^{-1} = \kappa \big( {\bf I} - ({\bf I} - \kappa {\bf X}) \big)^{-1} = \kappa \sum_{\ell = 0 }^{\infty} ({\bf I} - \kappa {\bf X})^{\ell}
\end{equation}
where the second equality holds if the parameter $\kappa$ is selected such that $0 < \kappa < \frac{2}{\max_n \lambda_n({\bf X})}$.
\end{lemma}
\begin{IEEEproof}
The inverse of an Hermitian matrix can be computed by inverting each eigenvalue, while keeping the eigenvectors fixed. This lemma follows by applying the standard Taylor series expansion  $(1-x)^{-1} = \sum_{\ell = 0}^{\infty} x^{\ell}$, for any $|x|<1$, on each eigenvalue of the Hermitian matrix $({\bf I} - \kappa {\bf X})$. The condition on $x$ corresponds to requiring that the spectral norm $\| {\bf I} - \kappa {\bf X} \|_2$ is bounded by unity, which holds for $ \kappa < \frac{2}{\max_n \lambda_n({\bf X})}$. See \cite{Sessler2005a} for an in-depth analysis of such properties of polynomial expansions.
\end{IEEEproof}

This lemma shows that the inverse of any Hermitian matrix can be expressed as a matrix polynomial. More importantly, the low-order terms are the most influential ones, since the eigenvalues of $({\bf I} - \kappa {\bf X})^{\ell}$ converge geometrically to zero as $\ell$ grows large. This is due to each eigenvalue $\lambda$ of $({\bf I} - \kappa {\bf X})$ having an absolute value smaller than unity, $|\lambda|<1$, and thus $\lambda^{\ell}$ goes geometrically to zero as $\ell \rightarrow \infty$. As such, it makes sense to consider a TPE of the matrix inverse using only the first $J$ terms.
This corresponds to approximating the inversion of each eigenvalue by a Taylor polynomial with $J$ terms, hence the approximation accuracy per matrix element is independent of $M$ and $K$; that is, $J$ needs not change with the system dimensions.

TPE has been successfully applied for low-complexity multi-user detection in \cite{Moshavi1996a,Honig2001a,Sessler2005a,hoydis} and channel estimation in \cite{Shariati2013a}. Next, we exploit the TPE technique to approximate RZF precoding by a matrix polynomial. Starting from $\bG_{\rm RZF}$ in \eqref{eq:RZF}, we note that
\begin{align} \label{eq:RZF-original}
	\beta  &\left(\frac{1}{K}\widehat{\bH}\wbHh +\xi {\bf I}_M \right)^{-1}  \frac{\widehat{\bH}}{\sqrt{K}}{\bf P}^{\frac{1}{2}} \\ \label{eq:RZF-approx1}
	       &= \beta \kappa \sum_{\ell = 0 }^{\infty} \left( {\bf I}_M - \kappa \Big(\frac{1}{K}\widehat{\bH}\wbHh +\xi {\bf I}_M \Big)  \right)^{\ell}  \frac{\widehat{\bH}}{\sqrt{K}}{\bf P}^{\frac{1}{2}} \\ \label{eq:RZF-approx2}
	       &\approx \beta \kappa \sum_{\ell = 0 }^{J-1} \left( {\bf I}_M - \kappa \Big(\frac{1}{K}\widehat{\bH}\wbHh +\xi {\bf I}_M \Big)  \right)^{\ell}  \frac{\widehat{\bH}}{\sqrt{K}}{\bf P}^{\frac{1}{2}} \\ 
\label{eq:RZF-approx4}
& =  \sum_{\ell = 0 }^{J-1} \left( \beta \kappa \sum_{n=\ell}^{J-1} \binom{n}{\ell} (1-\kappa \xi)^{n-\ell} (-\kappa)^{\ell} \right) \nonumber\\
&\times \Big(\frac{1}{K}\widehat{\bH}\wbHh \Big)^{\ell}\frac{\widehat{\bf H}}{\sqrt{K}}{\bf P}^{\frac{1}{2}}
\end{align}
where \eqref{eq:RZF-approx1} follows directly from Lemma \ref{lemma:inversion-expansion} (for an appropriate selection of $\kappa$),
\eqref{eq:RZF-approx2} is achieved by truncating the polynomial (only keeping the first $J$ terms), and \eqref{eq:RZF-approx4} follows from applying the binomial theorem and gathering the terms for each exponent.
Inspecting \eqref{eq:RZF-approx4}, we have a precoding matrix with the structure
\begin{equation}
	\bG_{\rm TPE}=\sum_{\ell=0}^{J-1} w_\ell \left(\frac{1}{K}\widehat{\bH} {\wbHh}\right)^{\ell}\frac{\widehat{\bH}}{\sqrt{K}}{\bf P}^{\frac{1}{2}}
\label{eq:precoder}
\end{equation}
where $w_0,\ldots,w_{J-1}$ are scalar coefficients. Although the bracketed term in \eqref{eq:RZF-approx4} provides a potential expression for $w_\ell$, we stress that these are generally not the optimal coefficients when $J < \infty$. Also, these coefficients are not satisfying the power constraint in \eqref{eq:power-constraint} since the coefficients are not adapted to the truncation.
Hence, we treat $w_0,\ldots,w_{J-1}$ as design parameters that should be selected to maximize the performance; for example, by maximizing the limiting value of the SINRs, as was done in Theorem \ref{theorem:optimal-regularization} for RZF precoding.
We note especially  that the value of $\kappa$ in \eqref{eq:RZF-approx4} does not need to be explicitly known in order to choose, optimize and implement the coefficients. We only need for $\kappa$ to exist, which is always the case under Assumption~A-\ref{eq:channel-model}.
Besides the simplified structure, the proposed precoding matrix $\bG_{\rm TPE}$ possesses a higher number of degrees of freedom (represented by the $J$ scalars $w_\ell$) than the RZF precoding (which has only the regularization coefficient $\xi$).

The precoding in \eqref{eq:precoder} is coined \emph{TPE precoding} and actually defines a whole class of precoding matrices for different $J$. For $J=1$ we obtain $\bG = \frac{w_0}{\sqrt{K}} \widehat{\bH}{\bf P}^{\frac{1}{2}}$, which equals MRT. Furthermore, RZF precoding can be obtained by choosing $J=\min(M,K)$ and coefficients based on the characteristic polynomial of $(\frac{1}{K}\widehat{\bH}\wbHh +\xi {\bf I}_M )^{-1}$ (directly from Cayley-Hamilton theorem). We refer to $J$ as the \emph{TPE order} and note that the corresponding polynomial degree is $J-1$. Clearly, proper selection of $J$ enables a smooth transition between the traditional low-complexity MRT and the high-complexity RZF precoding. Based on the discussion that followed Lemma \ref{lemma:inversion-expansion}, we assume that the parameter $J$ is a finite constant that does not grow with $M$ and $K$.

\section{Complexity Analysis}
\label{section:complexity-analysis}

In this section we compare the complexities of RZF and TPE precoding in a theoretical fashion and in an implementation sense.
The complexities are given as simple numbers of complex addition and multiplication operations needed for a given arithmetic operation. The number of floating point operations (flops) needed to implement these complex operations varies greatly according to the used hardware and complex number representation (i.e., polar or Cartesian). Thus, we will not attempt to give a measure in flops.
Also, the ability to parallelize operations and to customize algorithm-specific circuits has a fundamental impact on the computational delays and energy consumption in practical systems.

\subsection{Sum Complexity per Coherence Period for RZF and TPE}
\label{ssec:SumCompperCohTime}

In order to compare the number of complex operations needed for conventional RZF precoding and the proposed TPE precoding, it is important to consider how often each operation is repeated. There are two time scales: 1) operations that take place once per coherence period (i.e., once per channel realization) and 2) operations that take place every time the channel is used for downlink transmission. 
To differentiate between these time scales, we let $T_{\rm data}^{\rm pcp}$ denote the number of downlink channel uses for data transmission per coherence period. Recall from \eqref{eq:precoder_fun} that the transmit signal is $\bG {\bf s}$, where the precoding matrix $\bG \in \C^{M \times K}$ changes once per coherence period and the data transmit symbols ${\bf s} \in \C^{K \times 1}$ are different for each channel use.

The RZF precoding matrix in \eqref{eq:RZF} is computed once per coherence period. There are two equivalent expressions in \eqref{eq:RZF}, where the difference is that the matrix inversion is either of dimension $K \times K$ or $M \times M$. Since $K \leq M$ in most cases of practical interest, and especially in the massive MIMO regime, we consider the first precoding expression: $\frac{1}{\sqrt{K}} \widehat{\bH}\big(\frac{1}{\sqrt{K}}{\wbHh}\frac{1}{\sqrt{K}}\widehat{\bH}+\xi {\bf I}_K\big)^{-1}{\bf P}^{\frac{1}{2}}\beta$.

Assuming that $\frac{1}{\sqrt{K}} \widehat{\bH}$, $\xi$, $\beta$ and ${\bf P}^{\frac{1}{2}}$ are available in advance and the Hermitian operation is ``free'', we need to 1) compute the matrix-matrix multiplication $(\frac{1}{\sqrt{K}} {\wbHh}) (\frac{1}{\sqrt{K}}\widehat{\bH})$; 2) add the diagonal matrix $\xi {\bf I}_K$ to the result; 3) compute $\frac{1}{\sqrt{K}} \widehat{\bH}\big(\frac{1}{K}{\wbHh}\widehat{\bH}+\xi {\bf I}_K\big)^{-1}$; and 4) multiply the result with the diagonal matrix resulting from  ${\bf P}^{\frac{1}{2}} \beta$. These are standard operations for matrices, thus we obtain the numbers of complex operations as: $K^2 (2M-1)$, $K$, $\frac{K^3}{3}+2 K^2 M$, and $MK+K$ operations, respectively. 
Step 3) is not immediately obvious, but an efficient method for this part is to compute a Cholesky factorization of $\frac{1}{K}{\wbHh}\widehat{\bH}+\xi {\bf I}_K$ (at a cost of $K^3/3$) and then solve a simple linear equation system for each row of $\frac{1}{\sqrt{K}}{\wbHh}$ (at a cost of $2 K^2$ each) \cite[Slides 9-6, 9]{Boyd2008a}. This approach is  preferable to the alternative of completely inverting the matrix (again using Cholesky factorization) and then using matrix-matrix multiplication, as long as $K^3-KM > 0$. Given that the alternative method has a cost of $ 4K^3/3 + MK(2K-1) $. It is interesting to note here that, for the case of $M\gg K$, the matrix-matrix multiplication is actually more expensive than the matrix inversion ($2MK^2$ vs.\ $K^3$).
\footnote{Matrix multiplication combined with matrix inversion can be implemented using the Strassen's algorithm in \cite{Strassen1969a} and the improved Coppersmith-Winograd algorithm in \cite{Williams2012a}. These are divide-and-conquer algorithms that exploit that $2 \times 2$ matrices can be multiplied efficiently and thereby reduce the asymptotic complexity of multipling/inverting $K \times K$ matrices to $\mathcal{O}(K^{2.8074})$ and $\mathcal{O}(K^{2.373})$, respectively. Unfortunately, the overhead in these algorithms is heavy and thus $K$ needs to be at the order of several thousands to achieve a lower complexity than the Cholesky approach considered here. Hence, these alternative algorithms are unfavorable for matrices of practical sizes. }

Once $\bG_{\rm RZF}$ has been computed, the matrix-vector multiplication $\bG_{\rm RZF} {\bf s}$ requires $M(2 K-1)$ operations per channel use of data transmission. In summary, RZF precoding has a total number of complex operations per coherence period of
\begin{align*}
C_{\rm RZF}^{\rm pcp} = 4 K^2 M + \frac{ K^3}{3} + K(M+2) - K^2  +  T_{\rm data}^{\rm pcp} (2MK-M).
\end{align*}

\begin{figure}
	\centering
	\begin{tikzpicture}[scale=0.9,font=\normalsize]
       \renewcommand{\axisdefaulttryminticks}{4}
       \tikzstyle{every major grid}+=[style=densely dashed]
       \tikzstyle{every axis y label}+=[yshift=-10pt, xshift=1cm]
       \tikzstyle{every axis x label}+=[yshift=5pt]
       \tikzstyle{every axis legend}+=[cells={anchor=west},fill=white,
           at={(0.02,0.98)}, anchor=north west, font=\normalsize ]
       \begin{axis}[
%         ylabel absolute,
         ylabel style={yshift=-0.5cm},
         xmin=400,
         ymin=0,
         xmax=900,
           ymax=80000000,
         grid=major,
         scaled ticks=true,
      			xlabel={Coherence Period, $T_{\rm coherence}$},
      			ylabel={Complex Operations},
      	x post scale=1.2		
         ]
       \addplot[color=red, dashed, no marks, mark size=1.5pt,mark=x,line width=2pt] coordinates{
       (402,30473033) (410,30871033) (430,31866033) (450,32861033) (470,33856033) (490,34851033) (510,35846033) (530,36841033) (550,37836033) (570,38831033) (590,39826033) (610,40821033) (630,41816033) (650,42811033) (670,43806033) (690,44801033) (710,45796033) (730,46791033) (750,47786033) (770,48781033) (790,49776033) (810,50771033) (830,51766033) (850,52761033) (870,53756033) (890,54751033) (910,55746033) 
       };
       \addlegendentry{ {RZF} }
       \addplot[color=green, dashed, no marks, mark size=1.5pt,mark=x,line width=2pt] coordinates{
       (402,11522633) (410,12319033) (430,14310033) (450,16301033) (470,18292033) (490,20283033) (510,22274033) (530,24265033) (550,26256033) (570,28247033) (590,30238033) (610,32229033) (630,34220033) (650,36211033) (670,38202033) (690,40193033) (710,42184033) (730,44175033) (750,46166033) (770,48157033) (790,50148033) (810,52139033) (830,54130033) (850,56121033) (870,58112033) (890,60103033) (910,62094033) 
       };
       \addlegendentry{ {RZF2} }       
       \addplot[color=blue, no marks, mark size=2.5pt,mark=x,line width=1pt] coordinates{
       (402,100100) (410,500500) (430,1501500) (450,2502500) (470,3503500) (490,4504500) (510,5505500) (530,6506500) (550,7507500) (570,8508500) (590,9509500) (610,10510500) (630,11511500) (650,12512500) (670,13513500) (690,14514500) (710,15515500) (730,16516500) (750,17517500) (770,18518500) (790,19519500) (810,20520500) (830,21521500) (850,22522500) (870,23523500) (890,24524500) (910,25525500) 
       };
       \addlegendentry{ {TPE} }		
       \node[below] at (axis cs:800,20000000){\small $J=1$};
       \addplot[color=blue, no marks, mark size=2.5pt,mark=x,line width=1pt] coordinates{
       (402,300500) (410,1502500) (430,4507500) (450,7512500) (470,10517500) (490,13522500) (510,16527500) (530,19532500) (550,22537500) (570,25542500) (590,28547500) (610,31552500) (630,34557500) (650,37562500) (670,40567500) (690,43572500) (710,46577500) (730,49582500) (750,52587500) (770,55592500) (790,58597500) (810,61602500) (830,64607500) (850,67612500) (870,70617500) (890,73622500) (910,76627500) 
       };     
       \node[above] at (axis cs:750,55000000){\small $J=2$};
       \addplot[color=blue, no marks, mark size=2.5pt,mark=x,line width=1pt] coordinates{
       (402,500900) (410,2504500) (430,7513500) (450,12522500) (470,17531500) (490,22540500) (510,27549500) (530,32558500) (550,37567500) (570,42576500) (590,47585500) (610,52594500) (630,57603500) (650,62612500) (670,67621500) (690,72630500) (710,77639500) (730,82648500) (750,87657500) (770,92666500) (790,97675500) (810,102684500) (830,107693500) (850,112702500) (870,117711500) (890,122720500) (910,127729500) 
       }; 
       \node[above] at (axis cs:680,60000000){\small $J=3$};       
       \addplot[color=blue, no marks, mark size=2.5pt,mark=x,line width=1pt] coordinates{
       (402,701300) (410,3506500) (430,10519500) (450,17532500) (470,24545500) (490,31558500) (510,38571500) (530,45584500) (550,52597500) (570,59610500) (590,66623500) (610,73636500) (630,80649500) (650,87662500) (670,94675500) (690,101688500) (710,108701500) (730,115714500) (750,122727500) (770,129740500) (790,136753500) (810,143766500) (830,150779500) (850,157792500) (870,164805500) (890,171818500) (910,178831500) 
       }; 
       \node[above] at (axis cs:620,65000000){\small $J=4$};       
       \addplot[color=blue, no marks, mark size=2.5pt,mark=x,line width=1pt] coordinates{
       (402,901700) (410,4508500) (430,13525500) (450,22542500) (470,31559500) (490,40576500) (510,49593500) (530,58610500) (550,67627500) (570,76644500) (590,85661500) (610,94678500) (630,103695500) (650,112712500) (670,121729500) (690,130746500) (710,139763500) (730,148780500) (750,157797500) (770,166814500) (790,175831500) (810,184848500) (830,193865500) (850,202882500) (870,211899500) (890,220916500) (910,229933500) 
       }; 
       \node[above] at (axis cs:560,70000000){\small $J=5$};       
       \end{axis}
     \end{tikzpicture}
	\vskip-4mm
	\caption{Total number of arithmetic operations of RZF precoding and TPE precoding (with different $J$) for $K=100$ users and $M = 500$.} \label{figure-complexity2} 
	\vskip-4mm
\end{figure}
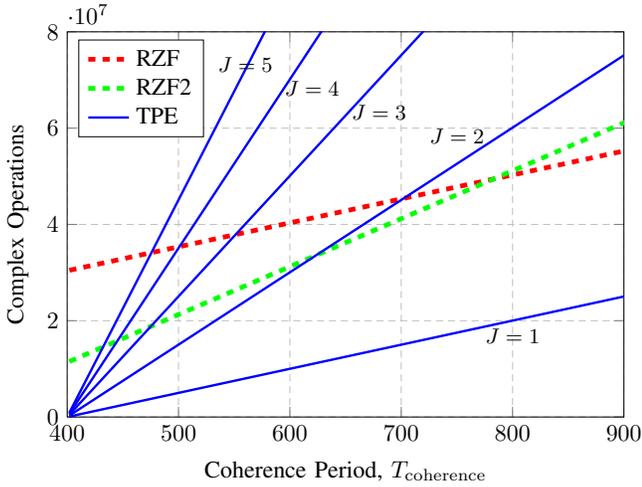

There is a second approach to looking at the RZF precoder complexity. Let the transmit signal with RZF precoding at channel use $t$ be denoted ${\bx}^{(t)}_{\rm RZF} $. The transmitted signal is then ${\bx}^{(t)}_{\rm RZF} = \bG_{\rm RZF} {\bf s}^{(t)}= \frac{1}{\sqrt{K}} \widehat{\bH}\big(\frac{1}{K}{\wbHh}\widehat{\bH}+\xi {\bf I}_K\big)^{-1}\beta{\bf P}^{\frac{1}{2}}{\bf s}^{(t)}$.
Thus, one can replace the ``matrix times inverse of another matrix'' operation taking place each coherence period, by a matrix-inverse operation per coherence period and two matrix-vector multiplications per data symbol vector. Thus, one effectively splits the previous point 3) in two parts and waits for the symbol vector to allow for the matrix-vector multiplications. This results in
\begin{equation*}
C_{\rm RZF2}^{\rm pcp} = 2 K^2 M + \frac{ 4K^3}{3} - K^2 +2K +  T_{\rm data}^{\rm pcp} (4MK-2M+K).
\end{equation*}
Still, this complexity is dominated by the matrix-matrix multiplication inside the inverse. However, the per coherence period complexity is reduced in exchange for a slight increase in complexity per symbol.
Depending on the use-case of the precoder, this change can either be advantageous or disadvantageous (see Figure~\ref{figure-complexity2} and Subsection~\ref{ssec:Delay1st}).
We note that choosing to incorporate the multiplication with ${\bf P}^{\frac{1}{2}}$ per coherence period or per symbol vector does only insignificantly change the stated outcomes.
In the following we will chose the appropriate version for each comparison. 

Next, we consider TPE precoding. Similar to before, we assume that $\frac{1}{\sqrt{K}} \widehat{\bH}$, $w_\ell$ and ${\bf P}^{\frac{1}{2}}$ are available in advance and the Hermitian operation is ``free''. Let the transmit signal vector with TPE precoding at channel use $t$ be denoted ${\bx}^{(t)}_{\rm TPE} $ and observe that it can be expressed as
\begin{align*}
{\bx}^{(t)}_{\rm TPE} = \bG_{\rm TPE} {\bf s}^{(t)} = \sum_{\ell=0}^{J-1} w_\ell \tilde{{\bx}}_{\ell}^{(t)}
\end{align*}
where ${\bf s}^{(t)}$ is the vector of data symbols at channel use $t$ and
\begin{equation*}
\tilde{{\bx}}_{\ell}^{(t)} =
	\begin{cases} 	
		\frac{\widehat{\bH}}{\sqrt{K}} ( {\bf P}^{\frac{1}{2}} {\bf s}^{(t)}), & \ell=0, \\
		\frac{\widehat{\bH}}{\sqrt{K}} (\frac{\wbHh}{\sqrt{K}} \tilde{{\bx}}_{\ell-1}^{(t)} ), & 1\leq \ell \leq J-1.
	\end{cases}
\end{equation*}
This reveals that there is an iterative way of computing the $J$ terms in TPE precoding. The benefit of this approach is that it can be implemented using only matrix-vector multiplications.\footnote{Intuitively one circumvents the expensive matrix-matrix multiplication with a domino-like chain of $2J-1$ (less expensive) matrix-vector multiplications per transmitted symbol vector. This became possible by replacing the inverse of a matrix-matrix multiplication in the RZF with a sum of weighted matrix powers.} 

Similar to above, we conclude that the case $\ell = 0$ uses $K + M(2K-1)$ operations and each of the $J-1$ cases of $\ell \geq 1$ needs $M(2K-1)+ K(2M-1)$ operations. One remarks that it is impractical and unneeded to carry out a matrix-matrix multiplication at this step. Finally, the multiplication with $w_\ell$ and the summation requires $M (2J-1)$ further operations. In summary, TPE precoding has a total number of arithmetic operations of
\begin{align*}
C_{\rm TPE}^{\rm pcp} = T_{\rm data}^{\rm pcp} \big( (4J-2)MK + (J-1)M + K(2-J) \big).
\end{align*}

When comparing RZF and TPE precoding, we note that the complexity of precomputing the RZF precoding matrix is very large, but it is only done once per coherence period. The corresponding matrix $\bG_{\rm TPE}$ for TPE precoding is never computed separately, but only indirectly as $\bG_{\rm TPE} {\bf s}$ for each data symbol vector ${\bf s}$. Intuitively, precomputation is beneficial when the coherence period is long (compared to $M$ and $K$) and the sequential computation of TPE precoding is beneficial when the system dimensions $M$ and $K$ are large (compared to the coherence period) or the coherence period is short. This is seen from the large dimensional complexity scaling which is $\mathcal{O}(4 K^2 M)$ or $\mathcal{O}(2 K^2 M)$ for RZF precoding (the latter, if the RZF or RZF2 approach is used) and $\mathcal{O}(4 J K M T_{\rm data}^{\rm pcp})$ for TPE precoding; thus, the asymptotic difference is significant. The break even point, where TPE precoding outperforms RZF is easily computed looking at $C_{\rm RZF}^{\rm pcp} > C_{\rm TPE}^{\rm pcp}$
\begin{align*}
 \Rightarrow T_{\rm data}^{pcp} <  \frac{ 4 K^2 M + \frac{K^3}{3} + K(M+2) - K^2 }{ 4(J-1)MK + JM + (2-J)K } \approx \frac{K}{J-1}
\end{align*}
and similar for $C_{\rm RZF2}^{\rm pcp} > C_{\rm TPE}^{\rm pcp}$.

One should not forget the overhead signaling required to obtain CSI at the UTs, which makes the number of channel uses $T_{\rm data}$ available for data symbols reduce with $K$. For example, suppose $T_{\rm coherence}$ is the total coherence period and that we use a TDD protocol, where $\eta_{\rm DL}$ is the fraction used for downlink transmission and $\mu K$ channel uses (for some $\mu \geq 1$) are consumed by downlink pilot signals that provide the UTs with sufficient CSI. We then have $T_{\rm data} = \eta_{\rm DL} T_{\rm coherence} - \mu K$. Using this relationship, the number of arithmetic operations are illustrated numerically in Fig.~\ref{figure-complexity2} for $\eta_{\rm DL}  = \frac{1}{2}$, $K = 100$, and $\mu = 2$.\footnote{These parameter values correspond to symmetric downlink/uplink transmission, 2 downlink pilot symbols per UT (at different frequencies). Looking at values similar the LTE standard \cite[Chapter 10]{Dahlman2011a}, e.g., a coherence bandwidth of $200$ kHz, and a coherence period of $5$ ms one would arrive a $T_{\rm coherence}$ of $1000$.} This figure shows that TPE precoding uses fewer operations than RZF precoding when the coherence period is short and the TPE order is small, while RZF is competitive for long coherence times.

We remark that all previously found results change in favor of TPE, if one uses the canonical transformation of complex to real operations by doubling all dimensions. 

\begin{remark}[Power Normalization]
In this section we assumed that $\beta$ and $w_\ell$ (and $\xi$) are known beforehand. These factors are responsible for the power normalization of the transmit signal. Depending on the chosen normalization, for example the average per UT one in this paper, it requires the full precoding matrix to be known. Thus it forbids the alternative implementation of RZF precoding detailed before. Note that this could be remedied by changing to ``strict'' per UT normalization.
In general, we can find values for $\beta$ and $w_\ell$, that only rely on channel statistics and are valid in the large-$(M,K)$ regime. This, and the possible fix for the alternative RZF approach, have motivated us to assume $\beta$ and $w_\ell$ as known.
\end{remark}

\subsection{Delay to First Transmission for RZF and TPE}
\label{ssec:Delay1st}
A practically important complexity metric is the number of complex operations for the first channel use. This number can also be interpreted as the delay until the start of data transmission.
This complexity can easily be found from the previous results, by choosing $T_{\rm data} = 1$. Directly looking at the massive MIMO case, we find
$C_{\rm RZF}^{\rm 1st} = 4 M K^2$, 
$C_{\rm RZF2}^{\rm 1st} = 2 M K^2$
and
$ C_{\rm TPE}^{\rm 1st} = 4JMK$.
Hence, the first data vector is transmitted by a factor of $K/(2J)$ earlier\footnote{Depending on the massive MIMO system $K$ can be on the order of $100$ and $M$ of the order $10K$, while we will see later that $J=4$ is sufficient for many cases.}, when TPE precoding is employed. This factor is significant and gives TPE precoding practical relevance, especially in massive MIMO systems and in very fast changing environments, i.e., when coherence periods are very short. We also remark that not wasting time during the coherence period pays off greatly, as the lost channel uses are given by the saved time multiplied by the (often large) coherence bandwidth.

\subsection{Implementation Complexity of RZF and TPE Precoding}
\label{subsec:implement-complexity}

\begin{figure}
\begin{center}
\includegraphics[width=\columnwidth]{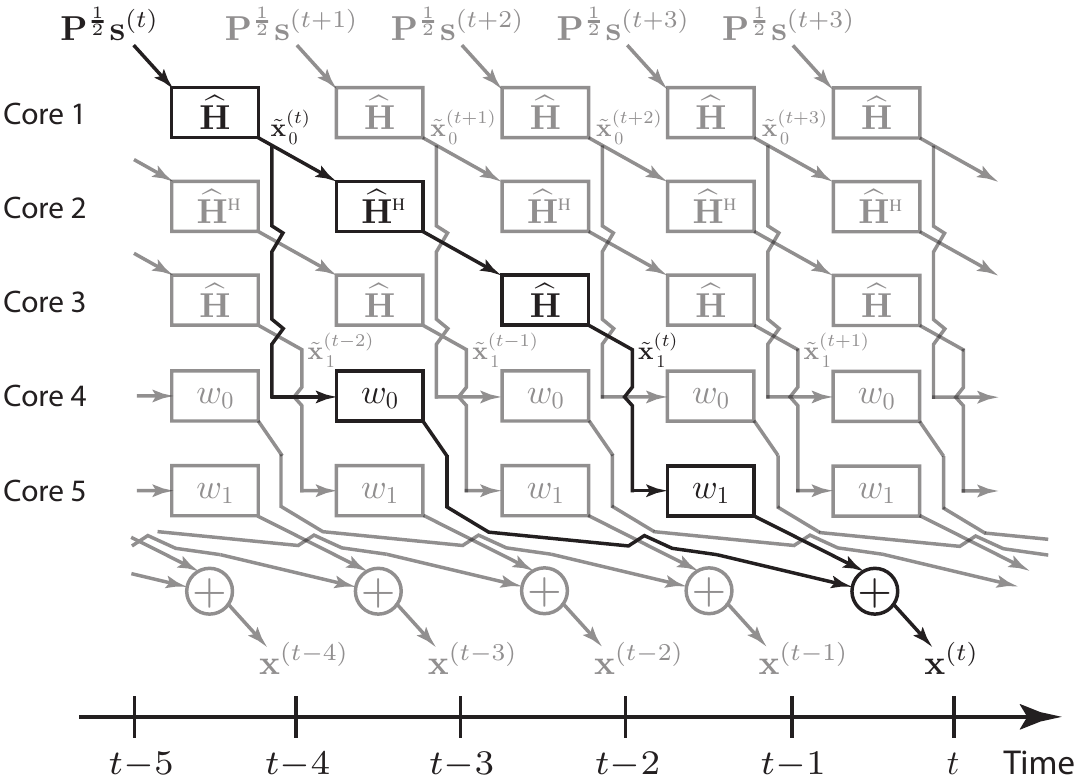}
\end{center} \vskip-5mm
\caption{Illustration of a simple pipelined implementation of the proposed TPE precoding with $J=2$, which removes the delays caused by precomputing the precoding matrix. Each block performs a simple matrix-vector multiplication, which enables highly efficient hardware implementation and $J$ can be increased by simply adding additional cores.} \label{figure-pipelining2} \vskip-4mm
\end{figure}

In practice, the number of arithmetic operations is not the main issue, but the implementation cost in terms of hardware complexity, time delays, and energy consumption. The analysis in Subsection~\ref{ssec:SumCompperCohTime} showed that we can only expect improvements in the sum of complex operations from TPE precoding per coherence period in certain scenarios. However, one advantage of TPE precoding is that it enables multistage hardware implementation where the computations are pipelined \cite{Sessler2005a} over multiple processing cores (e.g., application-specific integrated circuits (ASICs)). This structure is illustrated in Fig.~\ref{figure-pipelining2}, where the transmitted signal ${\bx}^{(t)}$ is prepared in the various cores (black path), while the preceding and succeeding transmit signals are computed in the ``free'' cores (grey paths). Each processing core performs two simple matrix-vector multiplications, each requiring approximately $\mathcal{O} (2MK)$ complex additions and multiplications per coherence period. This is relatively easy to implement using ASICs or FPGAs, which are know to be very energy-efficient and have low production cost. Consequently, we can select the TPE order $J$ as large as needed to obtain a certain precoding accuracy, if we are prepared to use as many circuits of the same type as needed. Then, the delay between two consecutive transmitted symbol vectors is given only by the delay of two matrix-vector multiplications.

In comparison, the inversion of RZF precoding can only be pseudo-parallelized by using tree structures.
Hence, the pipelining of the $C_{\rm RZF}$ complex operations per coherence period is limited by the delay of a single processing core that implements the inverse of a matrix-matrix; this delay is most probably much larger than the two  matrix-vector multiplications of TPE.
The delay of a second core implementing the multiplication of the inverse with the channel matrix is negligible in comparison.
Like mentioned before, the precomputation of the RZF precoding matrix causes non-negligible delays that forces $T_{\rm data}^{\rm pcp}$ to be smaller than for TPE precoding; 
for example, \cite{Shepard2012a} describes a hardware implementation from \cite{Dick2007a} where it takes $0.15$ ms to compute RZF precoding for $K=15$, which translated to a loss of $0.15\text{ms} \cdot 200\text{kHz} = 30$ channel uses in a system with coherence bandwidth $200$kHz.
Also, the number of active UTs can be much larger than this in large-scale MIMO systems \cite{Bjornson2013f}. 
TPE precoding does not cause such delays because there are no precomputations---the arithmetic operations are spread over the coherence period.

In practice this means one can argue that only the curve pertaining to $J=1$ in Fig.~\ref{figure-complexity2} is relevant for comparisons between TPE and RZF after implementation; if one is prepared to add (seemingly unfairly) as many computation cores as necessary to TPE.

\section{Analysis and Optimization of TPE Precoding}
In this section, we consider the large-$(M,K)$ regime, defined in Assumption \mbox{A-\ref{ass:regime}}. We show that SINR$_k$, for $k=1,\ldots,K$, under TPE precoding converges to a limit, a deterministic equivalent, that depends only on the coefficients $w_\ell$, the respective attributed power $p_k$, and the channel statistics.

Recall the SINR expression in \eqref{eq:sinr} and observe that ${\bf g}_k = \bG {\bf e}_k$ and $\bhh\bG_k\bG_k^{\mbox{\tiny H}}\bh_k =
\bhh\bG\bG^{\mbox{\tiny H}}\bh_k - \bhh{\bf g}_k{\bf g}_k^{\mbox{\tiny H}}\bh_k$, where ${\bf e}_k$ is the $k$th column of the identity matrix ${\bf I}_K$. By substituting the TPE precoding expression \eqref{eq:precoder} into \eqref{eq:sinr}, it is easy to show that the SINR writes as
\begin{equation}
	{\rm SINR}_k=\frac{{\bf w}^{\mbox{\tiny H}}{\bf A}_k{\bf w}}{{\bf w}^{\mbox{\tiny H}}{\bf B}_k{\bf w}+\sigma^2}
\label{eq:sinr_ob}
\end{equation}
where ${\bf w}=[w_0 \, \ldots \, w_{J-1}]^{\mbox{\tiny T}} $ and the $(\ell,m)$th elements of the matrices ${\bf A}_k$, ${\bf B}_k \in \mathbb{C}^{J \times J}$ are
\begin{align} \label{eq:A_k}
	\left[{\bf A}_k\right]_{\ell,m}&\!=\!\frac{p_k}{K}{\bf h}_k^{\mbox{\tiny H}}\left(\frac{1}{K}\widehat{\bf H}\widehat{\bf H}^{\mbox{\tiny H}}\right)^{\! \ell}\widehat{\bf h}_k \widehat{\bf h}_k^{\mbox{\tiny H}}\left(\frac{1}{K}\widehat{\bf H}\widehat{\bf H}^{\mbox{\tiny H}}\right)^{\! m} {\bf h}_k \\
	\left[{\bf B}_k\right]_{\ell,m}&\!=\!\frac{1}{K}{\bf h}_k^{\mbox{\tiny H}}\left(\frac{1}{K}\widehat{\bf H}\widehat{\bf H}^{\mbox{\tiny H}}\right)^{\!\ell}\widehat{\bf H}{\bf P}\widehat{\bf H}\left(\frac{1}{K}\widehat{\bf H}\widehat{\bf H}^{\mbox{\tiny H}}\right)^{\! m} {\bf h}_k -\left[{\bf A}_k\right]_{\ell,m} \label{eq:B_k}
\end{align}
for $\ell=0,\ldots,J-1$ and $m = 0,\ldots,J-1$.\footnote{The entries of matrices are numbered from $0$, for notational convenience.}

Since the random matrices ${\bf A}_k$ and ${\bf B}_k$ are of finite dimensions, it suffices to determine a deterministic equivalent for each of their elements. To achieve this, we express them using the resolvent matrix of $\widehat{\bf H}$. This can be done by introducing the following random functionals in $t$ and $u$:
\begin{align} \label{eq:X}
	& X_{k,M}(t,u)= \nonumber \\
	& \ \frac{1}{K^2}{\bf h}_k^{\mbox{\tiny H}} \Big(\frac{t}{K}\widehat{\bf H}\widehat{\bf H}^{\mbox{\tiny H}}+{\bf I}_M\Big)^{\!-1} \widehat{\bf h}_k\widehat{\bf h}_k^{\mbox{\tiny H}}\left(\frac{u}{K}\widehat{\bf H}\widehat{\bf H}^{\mbox{\tiny H}}+{\bf I}_M\right)^{\!-1}{\bf h}_k 
\\
	& Z_{k,M}(t,u)= \nonumber \\
	& \ \frac{1}{K}{\bf h}_k^{\mbox{\tiny H}} \Big(\frac{t}{K}\widehat{\bf H}\widehat{\bf H}^{\mbox{\tiny H}}+{\bf I}\Big)^{\!-1}{\widehat{\bf H}{\bf P}\widehat{\bf H}}\left(\frac{u}{K}\widehat{\bf H}\widehat{\bf H}^{\mbox{\tiny H}}+{\bf I}_K\right)^{\!-1} {\bf h}_k. \label{eq:Z}
\end{align}
By taking derivatives of $X_{k,M}(t,u)$ and $Z_{k,M}(t,u)$, we obtain
\begin{align} 	
X_{k,M}^{(\ell,m)} &= \frac{(-1)^{\ell+m} \ell! m!}{K^2}{\bf h}_k^{\mbox{\tiny H}} \left(\frac{\widehat{\bf H}\widehat{\bf H}^{\mbox{\tiny H}}}{K}\right)^\ell \widehat{\bf h}_k \widehat{\bf h}_k^{\mbox{\tiny H}}\left(\frac{\widehat{\bf H}\widehat{\bf H}^{\mbox{\tiny H}}}{K}\right)^m{\bf h}_k \label{eq:X-derivative}
\\
Z_{k,M}^{(\ell,m)} &= \frac{(-1)^{\ell+m} \ell!m!}{K} \, {\bf h}_k^{\mbox{\tiny H}} \left(\frac{\widehat{\bf H}\widehat{{\bf H}^{\mbox{\tiny H}}}}{K}\right)^\ell {\widehat{\bf H}{\bf P}\widehat{\bf H}} \left(\frac{\widehat{\bf H}\widehat{\bf H}^{\mbox{\tiny H}}}{K}\right)^m{\bf h}_k. \label{eq:Z-derivative}
\end{align}
Substituting \eqref{eq:X-derivative}--\eqref{eq:Z-derivative} into \eqref{eq:A_k}--\eqref{eq:B_k}, we obtain the alternative expressions
\begin{align*}
	\left[{\bf A}_k\right]_{\ell,m}
	&=\frac{K p_k(-1)^{\ell+m}}{\ell! m!}X_{k,M}^{(\ell,m)}\\	
	\left[{\bf B}_k\right]_{\ell,m}
	&=\frac{(-1)^{\ell+m}}{\ell! m!}(-Kp_kX_{k,M}^{(\ell,m)}+Z_{k,M}^{(\ell,m)}).	
\end{align*}
It, thus, suffices to study the asymptotic convergence of the bivariate functions $X_{k,M}(t,u)$ and $Z_{k,M}(t,u)$. This is achieved by the following new theorem and its corollary:
\begin{theorem}
	\label{th:x_k_z}
	Consider a channel matrix $\widehat{\bf H}$ whose columns are distributed according to Assumption A-\ref{ass:channel_imperfect}. Under the asymptotic regime described in Assumption A-\ref{ass:regime}, we have
	$$
	X_{k,M}(t,u)-\overline{X}_M(t,u)\xrightarrow[M,K\to+\infty]{\mathrm{a.s.}} 0
	$$
	and
	$$
	-Kp_kX_{k,M}(t,u)+Z_{k,M}(t,u) -\tr({\bf P}) \,\overline{b}_M(t,u) \xrightarrow[M,K\to+\infty]{\mathrm{a.s.}} 0
	$$
	where
\begin{align*}
	\overline{X}_M(t,u)&=\frac{(1-\tau^2)\delta(t)\delta(u)}{(1+t\delta(t))(1+u\delta(u))} \\
	\overline{b}_M(t,u)&=\left(\tau^2+\frac{(1-\tau^2)}{(1+u\delta(u))(1+t\delta(t))}\right) \beta_M(t,u) 		
\end{align*}
and $\beta_M(t,u)$ is given by
\begin{equation}
\label{eq:beta_M}
\beta_M(t,u)=\frac{\frac{1}{K}\tr\left(\bPhi{\bf T}(u)\bPhi{\bf T}(t) \right)}{(1+t\delta(t))(1+u\delta(u))-\frac{tu}{K}\tr \left( \bPhi{\bf T}(u)\bPhi{\bf T}(t) \right)}.
\end{equation}
\end{theorem}
\begin{IEEEproof}
	
	See Appendix \ref{app:x_k_z}.
\end{IEEEproof}
\begin{corollary}
	\label{corollary:derivative}
	Assume that Assumptions A-\ref{ass:channel} and A-\ref{ass:regime} hold true. Then, we have
	$$
 X_{k,M}^{(\ell,m)}-\overline{X}_{M}^{(\ell,m)}\xrightarrow[M,K\to+\infty]{\mathrm{a.s.}} 0
	$$
	and
	\begin{align*}
	\left( -Kp_kX_{k,M}^{(\ell,m)}+Z_{k,M}^{(\ell,m)}\right)-\tr({\bf P}) \, \overline{b}_{M}^{(\ell,m)}\xrightarrow[M,K\to+\infty]{\mathrm{a.s.}} 0.
	\end{align*}
\end{corollary}
\begin{IEEEproof}
	
	See Appendix~\ref{app:derivative}.
\end{IEEEproof}
Corollary~\ref{corollary:derivative} shows that the entries of ${\bf A}_k$ and ${\bf B}_k$, which depend on the derivatives of $X_{k,M}(t,u)$ and $Z_{k,M}(t,u)$, can be approximated in the asymptotic regime by  ${\bf T}^{(\ell)}$ and $\delta^{(\ell)}$, which are the derivatives of ${\bf T}(t)$ and $\delta(t)$ at $t=0$. Such derivatives can be computed numerically using the iterative algorithm of \cite{hoydis}, which is provided in Appendix~\ref{app:IterAlgoT} for the sake of completeness.

It remains to compute the aforementioned derivatives. To this end, we denote $f(t)=-\frac{1}{1+t\delta(t)}$, $\Tau(t)=-f(t){\bf T}(t)$ and by $f^{(\ell)}$, $\Tau^{(\ell)}$ their derivatives at $t=0$.  $\Tau^{(\ell)}$ can be calculated using the Leibniz derivation rule $ \Tau^{(\ell)} = (-{\bf T}(t) f(t) )^{(\ell)}|_{t=0} = - \sum_{n=0}^{\ell} {\ell\choose n} {\bf T}^{(n)} f^{(\ell-n)} $ and the respective values from Appendix~\ref{app:IterAlgoT}.
Rewriting \eqref{eq:beta_M} as
$$
\beta_M(t,u)\left(1-\frac{tu}{K}\tr \left( \bPhi\Tau(u)\bPhi\Tau(t) \right) \right)=\frac{1}{K}\tr \left( \bPhi\Tau(u)\bPhi\Tau(t) \right)
$$
and using the Leibniz rule, we obtain for any integers $\ell$ and $m$ greater than $1$, the expression
\begin{align*}
&\beta_M^{(\ell,m)}=\frac{1}{K}\tr \left( \bPhi \Tau^{(\ell)}\bPhi\Tau^{(m)} \right) \\
&+\sum_{k=1}^{\ell}\sum_{n=1}^{m} kn {\ell\choose k}{m\choose n}\beta_{M}^{(k-1,n-1)}\frac{1}{K}\tr \left( \bPhi\Tau^{(\ell-k)}\bPhi\Tau^{(m-n)} \right).
\end{align*}
An iterative algorithm for the computation of $\beta_M^{(\ell,m)}$ is given in Appendix~\ref{app:alg_iterbeta}.
With these derivation results on hand, we are now in the position to determine the expressions for the derivatives of the quantities of interest, namely $\overline{X}_{k,m}(t,u)$ and $\overline{b}_M(t,u)$. Using again the Leibniz derivation rule, we obtain
\begin{align*}
	&\overline{X}_M^{(\ell,m)}\!=\!(1-\tau^2)\sum_{k=0}^\ell\sum_{n=0}^m {\ell \choose k}{m \choose n}\delta^{(k)}\delta^{(n)}f^{(\ell-k)}f^{(m-n)} \\
 &\overline{b}_{M}^{(\ell,m)}\!=\!\tau^2\beta^{(\ell,m)}\!+\!(1-\tau^2)\sum_{k=0}^\ell\sum_{n=0}^m \! {\ell \choose k} \! {m\choose n}\beta_{M}^{(\ell-k,m-n)}\\
 &	\qquad \times f^{(k)}f^{(n)}.
\end{align*}
Using these results in combination with Corollary~\ref{corollary:derivative}, we immediately obtain the asymptotic equivalents of ${\bf A}_k$ and ${\bf B}_k$:
\begin{corollary}
Let $\widetilde{\bf A}$ and $\widetilde{\bf B}$ be the $J\times J$ matrices, whose entries are
\begin{align*}
\left[\widetilde{\bf A}\right]_{\ell,m}&=\frac{(-1)^{\ell+m}\overline{X}_M^{(\ell,m)}}{\ell!m!} \\
\left[\widetilde{\bf B}_k\right]_{\ell,m}&=\frac{(-1)^{\ell+m}\overline{b}_M^{(\ell,m)}}{\ell!m!}.
\end{align*}
Then, in the asymptotic regime, for any $k\in1,\ldots, K$ we have
$$
\max\left(\|{\bf A}_k-Kp_k \widetilde{\bf A}\|, \|{\bf B}_k-\tr \left({\bf P}\right)\widetilde{\bf B}\|\right)\xrightarrow[M,K\to+\infty]{\mathrm{a.s.}}0.
$$
\end{corollary}

\subsection{Optimization of the Polynomial Coefficients} 
\label{subsec:optimize-SINR}

Next, we consider the optimization of the asymptotic SINRs with respect to the polynomial coefficients ${\bf w} = [w_0 \, \ldots w_{J-1}]^T$. Using results from the previous sections, a deterministic equivalent for the SINR of the $k$th UT is
$$
{\gamma}_k=\frac{Kp_k{\bf w}^{\mbox{\tiny T}}\widetilde{\bf A}{\bf w}}{\tr \left({\bf P}\right){\bf w}^{\mbox{\tiny T}}\widetilde{\bf B}{\bf w}+\sigma^2}.
$$
 The optimized TPE precoding should satisfy the power constraints in \eqref{eq:power-constraint}:
\begin{equation}
\tr \left( {\bG}_{\rm TPE}{\bG}_{\rm TPE}^{\mbox{\tiny H}} \right) = P
\label{eq:condition}
\end{equation}
or equivalently
\begin{equation}
{\bf w}^{\mbox{\tiny T}}{\bf C}{\bf w}=P
\label{eq:constraint}
\end{equation}
where the $(\ell,m)$th element of the $J \times J$ matrix ${\bf C}$ is
\begin{equation}
\left[{\bf C}\right]_{\ell,m}= \frac{1}{K}\tr\left(\left(\frac{\wbH\wbHh}{K}\right)^\ell\wbH{\bf P}\wbHh\left(\frac{\wbH\wbHh}{K}\right)^m\right). \label{eq:C_k}
\end{equation}
In order to make the optimization problem independent of the channel realizations, we replace the constraint in \eqref{eq:constraint} by a deterministic one, which depends only on the statistics of the channel.
To find a deterministic equivalent of the matrix ${\bf C}$, we introduce the random quantity 
\begin{align*}
& Y_{M}(t,u)= \\  &\qquad \frac{1}{K}\tr\left(\left(\frac{t}{K}\wbH\wbHh+{\bf I}\right)^{-1}\wbH{\bf P}\wbHh \left(\frac{u}{K}\wbH\wbHh+{\bf I}\right)^{-1}\right)
\end{align*}
whose derivatives $Y_M^{(\ell,m)}$ satisfy
$$
\left[{\bf C}\right]_{\ell,m}=\frac{(-1)^{\ell+m}Y_M^{(\ell,m)}}{\ell!m!} .
$$
Using the same method as 
for the matrices ${\bf A}$ and ${\bf B}$, we achieve the following result:
\begin{theorem}
Considering the setting of Theorem~\ref{th:x_k_z}, we have the following convergence results:
\begin{enumerate}
\item Let $c(t,u)=\frac{\frac{1}{K}\tr \left( \bPhi{\bf T}(u){\bf T}(t) \right)}{(1+t\delta(t))(1+u\delta(u))}(1+tu\beta(t,u))$, then
$$
Y_M(t,u)-\tr \left({\bf P}\right)c(t,u)\xrightarrow[M,K\to+\infty]{\mathrm{a.s.}}0.
$$
\item Denote by $c^{(\ell,m)}$ the $\ell$th and $m$th derivatives with respect to $t$ and $u$, respectively, then 
\begin{align*}
c^{(\ell,m)}&=\sum_{k=1}^\ell\sum_{n=1}^m kn{\ell \choose k}{m \choose n}\beta^{(n-1,k-1)}\\
&\times\frac{1}{K}\tr \left( \bPhi \Tau^{(\ell-k)}\Tau^{(m-n)} \right) +\frac{1}{K}\tr\left(\bPhi\Tau^{(m)}\Tau^{(\ell)}\right)
\end{align*}
\item Let $\widetilde{\bf C}$ be the $J\times J$ matrix with entries given by
$$
[\widetilde{\bf C}]_{\ell,m}=\frac{(-1)^{\ell+m}c^{(\ell,m)}}{\ell!m!}.
$$
Then, in the asymptotic regime
$$
\|{\bf C}-\tr \left({\bf P}\right)\widetilde{\bf C}\|\xrightarrow[M,K\to+\infty]{\mathrm{a.s.}}  0.
$$
\end{enumerate}
\label{th:constraint}
\end{theorem}
\begin{IEEEproof}
The proof relies on the same techniques as before, so provide only a sketch in Appendix~\ref{app:scketch_c}.
 \end{IEEEproof}

Based on Theorem~\ref{th:constraint}, we can consider the deterministic power constraint
\begin{equation}
\tr \left({\bf P}\right){\bf w}^{\mbox{\tiny T}}\widetilde{\bf C}{\bf w}=P
\label{eq:deter_constraint}
\end{equation}
which can be seen as an approximation of \eqref{eq:constraint}, in the sense that for any ${\bf w}$ satisfying \eqref{eq:deter_constraint}, we have
$$
{\bf w}^{\mbox{\tiny T}}{\bf C}{\bf w}-P\xrightarrow[M,K\to+\infty]{\mathrm{a.s.}}0.
$$

Now the maximization of the asymptotic SINR of UT $k$ amounts to solving the following optimization problem:
\begin{equation}
\begin{split}
\maximize{{\bf w}}\, & \quad \frac{Kp_k{\bf w}^{\mbox{\tiny H}}\widetilde{\bf A}{\bf w}}{\tr \left({\bf P}\right){\bf w}^{\mbox{\tiny H}}\widetilde{\bf B}{\bf w}+\sigma^2} \\
\mathrm{subject} \,\, \mathrm{to} & \quad \, \tr \left({\bf P}\right){\bf w}^{\mbox{\tiny H}}\widetilde{\bf C}{\bf w}=P . \label{eq:prob}
\end{split}
\end{equation}

The next theorem shows that the optimal solution, ${\bf w}_{\rm opt}$, to \eqref{eq:prob} admits a closed-form expression.
  \begin{theorem}
	  \label{th:optimization}
	  \label{th:optimal_weights}
	  Let ${\bf a}$ be a unit norm eigenvector corresponding to the maximum eigenvalue $\lambda_{\max}$ of
\begin{equation} \label{eq:matrix-in-quotient}
\left(\widetilde{\bf B}+\frac{\sigma^2}{P}\widetilde{\bf C}\right)^{-\frac{1}{2}}\widetilde{\bf A}\left(\widetilde{\bf B}+\frac{\sigma^2}{P}\widetilde{\bf C}\right)^{-\frac{1}{2}}.
\end{equation}
Then  the optimal value of the problem in \eqref{eq:prob} is achieved by
\begin{equation} \label{eq:w-optimal}
	  {\bf w}_{\rm opt}=\sqrt{\frac{P}{\alpha \tr \left( {\bf P}\right)}} \left(\widetilde{\bf B}+\frac{\sigma^2}{P}\widetilde{\bf C}\right)^{-\frac{1}{2}}{\bf a}
\end{equation}
where the scaling factor $\alpha$ is
\begin{equation} \label{eq:alpha-optimal}
	  \alpha= \left\| \widetilde{\bf C}^{\frac{1}{2}}\left(\widetilde{\bf B}+\frac{\sigma^2}{P}\widetilde{\bf C}\right)^{-\frac{1}{2}}{\bf a}  \right\|^2.
\end{equation}
Moreover, for the optimal coefficients, the asymptotic SINR for the $k$th UT is
\begin{equation} \label{eq:asymptotic-SINR}
\gamma_k=\frac{Kp_k\lambda_{\rm max}}{\tr \left({\bf P} \right)}.
\end{equation}
  \end{theorem}
  \begin{IEEEproof}
	The proof is given Appendix~\ref{app:optimization}.
  \end{IEEEproof}

The optimal polynomial coefficients for UT $k$ are given in \eqref{eq:w-optimal} of Theorem \ref{th:optimal_weights}. Interestingly, these coefficients are independent of the user index, thus we have indeed derived the jointly optimal coefficients. Furthermore, all users converge to the same deterministic SINR up to an UT-specific scaling factor $\frac{K p_k}{\gamma \tr ({\bf P}) }$.

\begin{remark}
The asymptotic SINR expressions in \eqref{eq:asymptotic-SINR} are only functions of the statistics and the power allocation $p_1,\ldots,p_K$. The power allocation can be optimized with respect to some system performance metric. For example, one can show that the asymptotic average achievable rate
$$
\frac{1}{K}\sum_{k=1}^K \log_2\left(1+\frac{Kp_k\lambda_{\rm max}}{\tr \left( {\bf P} \right) }\right)
$$
is maximized by a uniform power allocation $p_k \! =\! \frac{P}{K}$ for all $k$.
\end{remark}

\begin{remark}
Theorem \ref{th:optimal_weights} shows that the $J$ polynomial coefficients that jointly maximize the asymptotic SINRs can be computed using only the channel statistics and the channel estimation error. The optimal coefficients are then  given in closed form in \eqref{eq:w-optimal}. Numerical experiments show that the coefficients are very robust to underestimation of $\tau$ and robust to overestimation.
Hence, the main feature of Theorem~\ref{th:optimization} is that the TPE precoding coefficients can be computed beforehand, or at least be updated at the relatively slow rate of change of the channel statistics. Thus, the cost of the optimization step is negligible with respect to calculating the precoding itself. The performance of finite-dimensional large-scale MIMO systems is evaluated numerically in Section~\ref{sec:sim}.
\end{remark}
\begin{remark}
Finally, we remark that Assumption A-\ref{ass:regime} prevents us from directly analyzing the scenario where $K$ is fixed and $M \rightarrow \infty$, but we can infer the behavior of TPE precoding based on previous works. In particular, it is known that MRT is an asymptotically optimal precoding scheme in this scenario \cite{Rusek2013a}. We recall from Section~\ref{subsection:polynomial-expansion} that TPE precoding reduces to MRT for $J=1$. Hence, we expect the optimal coefficients to behaves as $w_{0} \neq 0$ and $w_{\ell}\rightarrow 0$ for $\ell \geq 1$ when $M \rightarrow \infty$. In other words, we can reduce $J$ as $M$ grows large and still keep a fixed performance gap to RZF precoding.
\end{remark}

\section{Simulation Results}
\label{sec:sim}
In this section, we compare the RZF precoding from \cite{PEE05} (which was restated in \eqref{eq:RZF}) with the proposed TPE precoding (defined in \eqref{eq:precoder}) by means of simulations. The purpose is to validate the performance of the proposed precoding scheme and illustrate some of its main properties. The performance measure is the average achievable rate $$ \displaystyle r=\frac{1}{K}\sum_{k=1}^{K}\mathbb{E}[ \log_2(1+\mathrm{SINR}_k) ]$$ of the UTs, where the expectation is taken with respect to different channel realizations and users.
In the simulations, we model the channel covariance matrix as
$$
[\boldsymbol{\Phi}]_{i,j} = \begin{cases}  a^{j-i}, & i \leq j, \\  (a^{i-j})^*, & i > j \end{cases}
$$
where $a$ is chosen to be $0.1$. This approach is known as the exponential correlation model \cite{Loyka2001a}. More involved  models could be chosen here, but would make it harder to evaluate the performance and function of TPE, while not offering more insight.  The sum power constraint
  $$  
  \tr \left( {\bf G}_{\rm RZF/TPE}{\bf G}_{\rm RZF/TPE}^{\mbox{\tiny H}} \right)=P
  $$
is applied for both precoding schemes. Unless otherwise stated, we use uniform power allocation for the UTs, since the asymptotic properties of RZF precoding are known in this case (see Theorem~\ref{theorem:optimal-regularization}).
Without loss of generality, we have set $\sigma^2=1$. 
Our default simulation model is a large-scale single-cell MIMO system of dimensions $M=128$ and $K=32$.

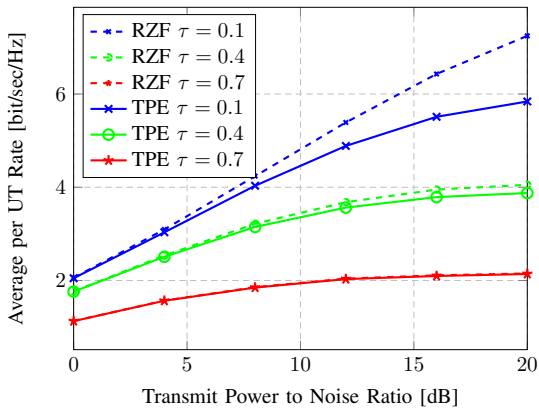
\begin{figure}
  \centering
   \begin{tikzpicture}[scale=0.8,font=\normalsize]
    \renewcommand{\axisdefaulttryminticks}{4}
    \tikzstyle{every major grid}+=[style=densely dashed]
    \tikzstyle{every axis y label}+=[yshift=-10pt]
    \tikzstyle{every axis x label}+=[yshift=5pt]
    \tikzstyle{every axis legend}+=[cells={anchor=west},fill=white,
        at={(0.02,0.98)}, anchor=north west, font=\normalsize ]
    \begin{axis}[
      xmin=0,
      y filter/.code={\pgfmathparse{#1*1.4427}},	
      xmax=20,
      xtick={0,5,...,20},
      grid=major,
      scaled ticks=true,
   			xlabel={Transmit Power to Noise Ratio [dB]},
   			ylabel={Average per UT Rate [bit/sec/Hz]},
   	x post scale=1.1,	
   	ylabel style={yshift=-0.3cm}   						
      ]
    \addplot[color=blue, dashed, mark size=1.5pt,mark=x,line width=1pt] coordinates{
    (0.000,1.427) (4.000,2.145) (8.000,2.938) (12.000,3.736) (16.000,4.457) (20.000,5.024) 
    };
    \addlegendentry{ {RZF $\tau=0.1$} }
    
    \addplot[color=green, dashed, mark size=1.5pt,mark=o,line width=1pt] coordinates{
    (0.000,1.225) (4.000,1.758) (8.000,2.230) (12.000,2.551) (16.000,2.738) (20.000,2.809) 
    };
    \addlegendentry{ {RZF $\tau=0.4$} }
    
    \addplot[color=red, dashed, mark size=1.5pt,mark=star,line width=1pt] coordinates{
    (0.000,0.783) (4.000,1.087) (8.000,1.287) (12.000,1.414) (16.000,1.463) (20.000,1.491) 
    };
    \addlegendentry{ {RZF $\tau=0.7$} }
    
    \addplot[color=blue ,mark size=2.5pt,mark=x,line width=1pt] coordinates{
    (0.000,1.420) (4.000,2.103) (8.000,2.795) (12.000,3.387) (16.000,3.820) (20.000,4.048) 
    };
    \addlegendentry{ {TPE $\tau=0.1$} }		
    
    \addplot[color=green ,mark size=2.5pt,mark=o,line width=1pt] coordinates{
    (0.000,1.221) (4.000,1.739) (8.000,2.182) (12.000,2.470) (16.000,2.626) (20.000,2.687) 
    };
    \addlegendentry{ {TPE $\tau=0.4$} }		
    
    \addplot[color=red ,mark size=2.5pt,mark=star,line width=1pt] coordinates{
    (0.000,0.782) (4.000,1.085) (8.000,1.281) (12.000,1.407) (16.000,1.453) (20.000,1.483) 
    };
    \addlegendentry{ {TPE $\tau=0.7$} }			
	  
    \end{axis}
  \end{tikzpicture} \vskip-3mm
  \caption{Average per UT rate vs. transmit power to noise ratio for varying CSI errors at the BS ($J=3$, $M=128$, $K=32$).}
  \label{fig:varytauL5} 
\end{figure}
We first take a look at Fig.~\ref{fig:varytauL5}. It considers a TPE order of $J=3$ and three different quality levels of the CSI at the BS: $\tau \in \{ 0.1,\, 0.4, \, 0.7 \}$. From Fig.~\ref{fig:varytauL5}, we see that RZF and TPE achieve almost the same average UT performance when a bad channel estimate is available ($\tau=0.7$). Furthermore, TPE and RZF perform almost identically at low SNR values, for any $\tau$. In general, the unsurprising observation is that the rate difference becomes larger at high SNRs and when $\tau$ is small (i.e., with more accurate channel knowledge).

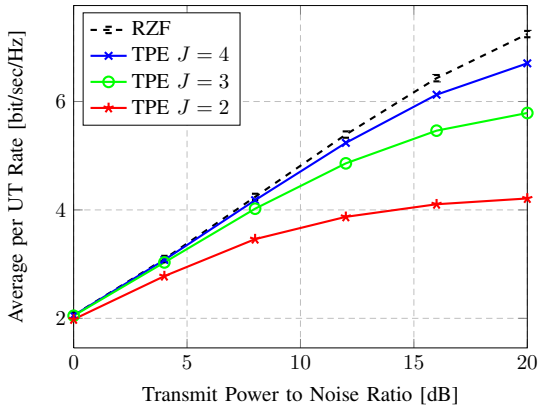
\begin{figure}
  \centering
   \begin{tikzpicture}[scale=0.8,font=\normalsize]
    \renewcommand{\axisdefaulttryminticks}{4}
    \tikzstyle{every major grid}+=[style=densely dashed]
    \tikzstyle{every axis y label}+=[yshift=-10pt]
    \tikzstyle{every axis x label}+=[yshift=5pt]
    \tikzstyle{every axis legend}+=[cells={anchor=west},fill=white,
        at={(0.02,0.98)}, anchor=north west, font=\normalsize ]
    \begin{axis}[
      xmin=0,
	  y filter/.code={\pgfmathparse{#1*1.4427}},	
      xmax=20,
      xtick={0,5,...,20},
      grid=major,
      scaled ticks=true,
   			xlabel={Transmit Power to Noise Ratio [dB]},
   			ylabel={Average per UT Rate [bit/sec/Hz]},
   	x post scale=1.1,	
   	ylabel style={yshift=-0.3cm}   						
      ]
	  \addplot[color=black, dashed, mark size=1.5pt,mark=square,line width=1pt] coordinates{	
	  (0.000,1.426) (4.000,2.145) (8.000,2.939) (12.000,3.736) (16.000,4.457) (20.000,5.021) 
	  };
      \addlegendentry{ {RZF} }
      
      \addplot[color=blue ,mark size=2.5pt,mark=x,line width=1pt] coordinates{
	  (0.000,1.424) (4.000,2.133) (8.000,2.897) (12.000,3.630) (16.000,4.246) (20.000,4.647) 
	  };
      \addlegendentry{ {TPE $J=4$} }			
   			
      \addplot[color=green ,mark size=2.5pt,mark=o,line width=1pt] coordinates{
	  (0.000,1.419) (4.000,2.101) (8.000,2.788) (12.000,3.368) (16.000,3.786) (20.000,4.012) 
	  };
      \addlegendentry{ {TPE $J=3$} }		
      
      \addplot[color=red ,mark size=2.5pt,mark=star,line width=1pt] coordinates{
	  (0.000,1.373) (4.000,1.926) (8.000,2.399) (12.000,2.684) (16.000,2.845) (20.000,2.919) 
	  };
      \addlegendentry{ {TPE $J=2$} }		
      
    \end{axis}
  \end{tikzpicture} \vskip-3mm
  \caption{Average UT rate vs. transmit power to noise ratio for different orders $J$ in the TPE precoding ($M=512$, $K=128$, $\tau=0.1$).}
  \label{fig:varyL} \vskip-3mm
\end{figure}
Fig.~\ref{fig:varyL} shows more directly the relationship between the average achievable UT rates and the TPE order $J$. We consider the case $\tau=0.1$, $M=512$, and $K=128$, in order to be in a regime where TPE performs relatively bad (see Fig.\ref{fig:varytauL5}) and the precoding complexity becomes an issue. 
From the figure, we see that choosing a larger value for $J$ gives a TPE performance closer to that of RZF. However, doing so will also require more hardware; see Section \ref{subsec:implement-complexity}. The proposed TPE precoding never surpasses the RZF performance, which is noteworthy since TPE has $J$ degrees of freedom that can be optimized (see Section \ref{subsec:optimize-SINR}), while RZF only has one design parameter. Hence one can regard RZF precoding as an upper bound to TPE precoding in the single-cell scenario.\footnote{The optimal precoding parametrization in \cite{Bjornson2012c} has $K-1$ parameters. To optimize some general performance metric, it is therefore necessary to let the number of design parameters scale with the system dimensions.}

\begin{figure}
  \centering
   \begin{tikzpicture}[scale=0.8,font=\normalsize]
    \renewcommand{\axisdefaulttryminticks}{4}
    \tikzstyle{every major grid}+=[style=densely dashed]
    \tikzstyle{every axis y label}+=[yshift=-10pt]
    \tikzstyle{every axis x label}+=[yshift=5pt]
    \tikzstyle{every axis legend}+=[cells={anchor=west},fill=white,
        at={(0.98,0.85)}, anchor=north east, font=\normalsize ]
    \begin{axis}[
      xmin=8,
        ymin=0,
      xmax=64,
        ymax=0.36,
      xtick={8,16,...,64},
	  y filter/.code={\pgfmathparse{#1*1.4427}},	
      grid=major,
      scaled ticks=true,
   			xlabel={Number of UTs, $K$},
   			ylabel style={align=center},
   			ylabel={Per UT Rate-Loss [bit/s/Hz]},
   	x post scale=1.1,	
   	ylabel style={yshift=-0.2cm}   						
      ]
      \addplot[color=blue ,mark size=2.5pt,mark=x,line width=1pt] coordinates{
      (8.000,0.209) (16.000,0.234) (24.000,0.225) (32.000,0.232) (40.000,0.238) (48.000,0.232) (56.000,0.244) (64.000,0.239) 
      };
      \addlegendentry{ {$J = 3$} }	
      
      \addplot[color=green, dashed ,mark size=2.5pt,mark=o,line width=1pt] coordinates{
      (8.000,0.099) (16.000,0.094) (24.000,0.093) (32.000,0.100) (40.000,0.097) (48.000,0.098) (56.000,0.102) (64.000,0.100) 
      };
      \addlegendentry{ {$J = 4$ ($12$dB)} }
      
      \addplot[color=green ,mark size=2.5pt,mark=o,line width=1pt] coordinates{
      (8.000,0.060) (16.000,0.063) (24.000,0.063) (32.000,0.062) (40.000,0.064) (48.000,0.062) (56.000,0.064) (64.000,0.060) 
      };
      \addlegendentry{ {$J = 4$} }		
      
      \addplot[color=red ,mark size=2.5pt,mark=star,line width=1pt] coordinates{
      (8.000,0.022) (16.000,0.016) (24.000,0.015) (32.000,0.013) (40.000,0.016) (48.000,0.016) (56.000,0.015) (64.000,0.014) 
      };
      \addlegendentry{ {$J = 5$} }			
      
    \end{axis}
  \end{tikzpicture} \vskip-4mm
  \caption{Rate-loss of TPE vs.~RZF with respect to growing $K$, where the ratio $M/K$ is fixed at $4$ and the average SNR is set to $10$ dB ($\tau=0.1$).}
  \label{fig:Figure3_RateLossvsK_EmilvaryMK} \vskip-2mm
\end{figure}
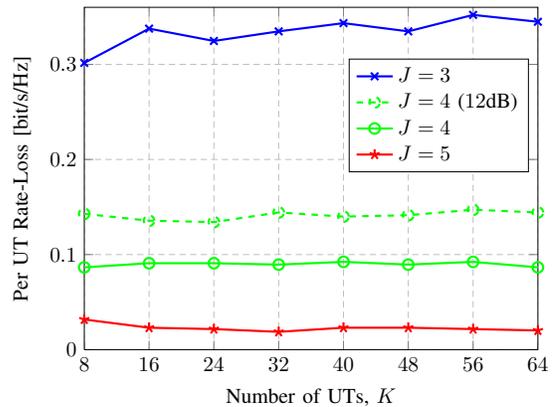
It is desirable to select the TPE order $J$ in such a way that we achieve a certain limited rate-loss with respect RZF precoding. Fig.~\ref{fig:Figure3_RateLossvsK_EmilvaryMK} illustrates the rate-loss (per UT) between TPE and RZF, while the number of UTs $K$ and transmit antennas $M$ increase with a fixed ratio ($M/K = 4$). The figure considers the case of $\tau=0.1$. We observe, that the TPE order $J$ and the system dimensions are independent in their respective effects on the rate-loss between TPE and RZF precoding. This observation is in line with previous results on polynomial expansions, for example \cite{Honig2001a} where reduced-rank received filtering was considered. The independence between $J$ and the system dimensions $M$ and $K$ (given the same ratio) is indeed a main motivation behind TPE precoding, because it implies that the order $J$ can be kept small even when TPE precoding is applied to very large-scale MIMO systems. The intuition behind this result is that the polynomial expansion approximates the inversion of each eigenvalue with the same accuracy, irrespective of the number of eigenvalues; see Section \ref{subsection:polynomial-expansion} for details.
Although the relative performance loss is unaffected by the system dimensions, we also see that $J$ needs to be increased along with the SNR, if a constant performance gap is desired.

\begin{figure}
  \centering
   \begin{tikzpicture}[scale=0.8,font=\normalsize]
    \renewcommand{\axisdefaulttryminticks}{4}
    \tikzstyle{every major grid}+=[style=densely dashed]
    \tikzstyle{every axis y label}+=[yshift=-10pt]
    \tikzstyle{every axis x label}+=[yshift=5pt]
    \tikzstyle{every axis legend}+=[cells={anchor=west},fill=white,
        at={(0.02,0.98)}, anchor=north west, font=\normalsize ]
    \begin{axis}[
      xmin=0,
	  y filter/.code={\pgfmathparse{#1*1.4427}},	
      xmax=20,
      xtick={0,5,...,20},
      ytick={2,3,...,4},
      grid=major,
      scaled ticks=true,
   			xlabel={Transmit Power to Noise Ratio [dB]},
   			ylabel={Average per UT Rate [bit/s/Hz]},
		x post scale=1.1,	
		ylabel style={yshift=-0.3cm}			
      ]
        \addplot[color=black, dashed, mark size=1.5pt,mark=square,line width=1pt] coordinates{	
        (0.000,1.222) (4.000,1.760) (8.000,2.239) (12.000,2.552) (16.000,2.723) (20.000,2.813) 
        };
        \addlegendentry{ {RZF} }
        			
        \addplot[color=blue ,mark size=2.5pt,mark=x,line width=1pt] coordinates{
        (0.000,1.233) (4.000,1.761) (8.000,2.221) (12.000,2.509) (16.000,2.665) (20.000,2.750) 
        };
        \addlegendentry{ {TPEopt} }
        			
        \addplot[color=green ,mark size=2.5pt,mark=o,line width=1pt] coordinates{
        (0.000,1.218) (4.000,1.741) (8.000,2.191) (12.000,2.466) (16.000,2.614) (20.000,2.695) 
        };
        \addlegendentry{ {TPE} }
      
    \end{axis}
  \end{tikzpicture} \vskip-4mm
  \caption{Average UT rate vs.~transmit power to noise ratio with RZF, TPE, and TPEopt precoding ($J=3$, $M=128$, $K=32$, $\tau=0.4$).}
  \label{fig:Figure4_AblaperfectOptimisation} \vskip-2mm
\end{figure}
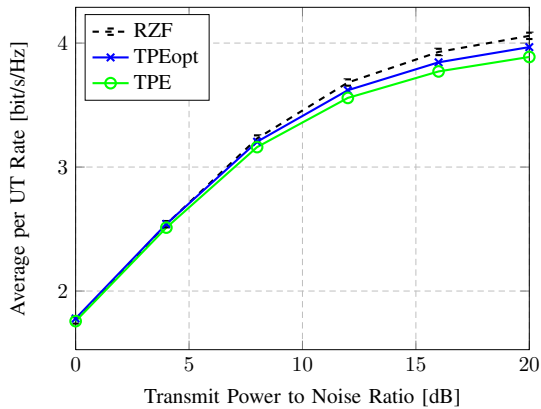
In the simulation depicted in Fig.~\ref{fig:Figure4_AblaperfectOptimisation}, we introduce a hypothetical case of TPE precoding (TPEopt) that optimizes the $J$ coefficients using the estimated channel coefficients in each coherence period, instead of relying solely on the channel statistics. More precisely, the optimal coefficients in Theorem~\ref{th:optimization} are not computed using the deterministic equivalents of $\widetilde{\bf A}$, $\widetilde{\bf B}$, and $\widetilde{\bf C}$, but using the original matrices from \eqref{eq:A_k}, \eqref{eq:B_k} and \eqref{eq:C_k}. This plot illustrates the additional performance loss caused by precalculating the TPE coefficients based on channel statistics and asymptotic analysis, instead of carrying out the optimization step for each channel realization. The difference is virtually zero at low SNRs and high at high SNRs. Furthermore, we note that increasing the value of $J$ has the same performance-gap-reducing effect on TPEopt, as it has on TPE (see Figs.~\ref{fig:varyL} and \ref{fig:Figure3_RateLossvsK_EmilvaryMK}). In order to preserve readability, only the curves pertaining to $J=3$ are shown in Fig.~\ref{fig:Figure4_AblaperfectOptimisation}.

\begin{figure}
  \centering
   \begin{tikzpicture}[scale=0.8,font=\normalsize]
    \renewcommand{\axisdefaulttryminticks}{4}
    \tikzstyle{every major grid}+=[style=densely dashed]
    \tikzstyle{every axis y label}+=[yshift=-10pt]
    \tikzstyle{every axis x label}+=[yshift=5pt]
    \tikzstyle{every axis legend}+=[cells={anchor=west},fill=white,
        at={(0.98,0.02)}, anchor=south east, font=\normalsize ]
    \begin{axis}[
    legend columns=4,
    column sep=5pt,
    transpose legend,
      xmin=0,
	  ymin=0.5,
	xmax=20,
        ymax=6.5,
      xtick={0,5,...,20},
      ytick={0,2,...,6},
      grid=major,
      scaled ticks=true,
   			xlabel={Transmit Power to Noise Ratio [dB]},
   			ylabel={Average Rate per UT Class [bit/s/Hz]},
   	x post scale=1.1,	
   	ylabel style={yshift=-0.3cm}		
      ]
	
      \addplot[color=black, no marks, mark size=1.5pt,mark=square,line width=1pt] coordinates{
     (0.000,2.586) (2.000,3.105) (4.000,3.636) (6.000,4.161) (8.000,4.661) (10.000,5.118) (12.000,5.515) (14.000,5.843) (16.000,6.098) (18.000,6.288) (20.000,6.422) 
     };
     \addlegendentry{ {DE $c_4$} }
     \addplot[color=black, dashed, only marks, mark options=solid, mark size=2.5pt,mark=o,line width=1pt] coordinates{
     (0.000,2.588) (2.000,3.113) (4.000,3.642) (6.000,4.177) (8.000,4.673) (10.000,5.145) (12.000,5.554) (14.000,5.877) (16.000,6.144) (18.000,6.339) (20.000,6.483) 
     };
     \addlegendentry{ {MC $c_4$} }
	
 	\addplot[color=blue, no marks, mark size=1.5pt,mark=square,line width=1pt] coordinates{	
    (0.000,2.249) (2.000,2.744) (4.000,3.259) (6.000,3.773) (8.000,4.265) (10.000,4.717) (12.000,5.111) (14.000,5.436) (16.000,5.690) (18.000,5.879) (20.000,6.012) 
    };
    \addlegendentry{ {DE $c_3$} }
    \addplot[color=blue, dashed, only marks, mark options=solid, mark size=3pt,mark=diamond,line width=1pt] coordinates{	
    (0.000,2.253) (2.000,2.751) (4.000,3.265) (6.000,3.784) (8.000,4.274) (10.000,4.735) (12.000,5.131) (14.000,5.474) (16.000,5.730) (18.000,5.933) (20.000,6.071) 
    };
    \addlegendentry{ {MC $c_3$} }
	
    \addplot[color=green, no marks ,mark size=2.5pt,mark=square,line width=1pt] coordinates{
    (0.000,1.808) (2.000,2.263) (4.000,2.747) (6.000,3.239) (8.000,3.717) (10.000,4.159) (12.000,4.546) (14.000,4.868) (16.000,5.119) (18.000,5.306) (20.000,5.439) 
    };
    \addlegendentry{ {DE $c_2$} }
    \addplot[color=green, dashed, only marks, mark options=solid, mark size=3pt,mark=star,line width=1pt] coordinates{
    (0.000,1.809) (2.000,2.269) (4.000,2.749) (6.000,3.250) (8.000,3.721) (10.000,4.176) (12.000,4.567) (14.000,4.892) (16.000,5.153) (18.000,5.346) (20.000,5.483) 
    };
    \addlegendentry{ {MC $c_2$} }
	
	\addplot[color=red, no marks, mark size=1.5pt,mark=square,line width=1pt] coordinates{	
	(0.000,1.171) (2.000,1.536) (4.000,1.948) (6.000,2.385) (8.000,2.823) (10.000,3.238) (12.000,3.607) (14.000,3.916) (16.000,4.160) (18.000,4.342) (20.000,4.471) 
	};
	\addlegendentry{ {DE $c_1$} }	
	\addplot[color=red, dashed, only marks, mark options=solid, mark size=2.5pt,mark=square,line width=1pt] coordinates{
	(0.000,1.172) (2.000,1.538) (4.000,1.950) (6.000,2.391) (8.000,2.828) (10.000,3.253) (12.000,3.631) (14.000,3.946) (16.000,4.200) (18.000,4.377) (20.000,4.507) 
	};
	\addlegendentry{ {MC $c_1$} }  	
  	
  \end{axis}
  \end{tikzpicture} \vskip-4mm
  \caption{Average rate per UT class vs.~transmit power to noise ratio with TPE precoding ($J=3$, $M=256$, $K=64$, $\tau=0.1$).}
  \label{fig:non_uniform}
\end{figure}
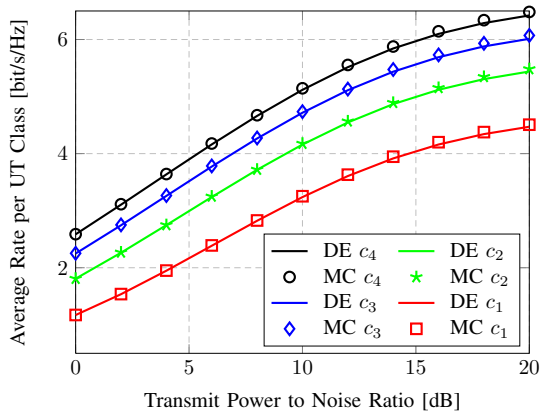
Finally, to assess the validity of our results, we treat the case of non-uniform power allocation (i.e., with different values for $p_k$). In particular, we considered a situation where the users are divided into four classes
corresponding to $\{c_1, c_2, c_4 c_4\}=\{1,2,3,4\}$, where $p_k = \frac{c_k}{K}$ in order to adhere to the scaling in Assumption A-\ref{ass:power}.
Fig.~\ref{fig:non_uniform} shows the theoretical large-($M,K$) regime ($\mathrm{DE}$; based on \eqref{eq:asymptotic-SINR}) and empirical ($\mathrm{MC}$; based on \eqref{eq:sinr_ob}) average rate per UT for each class, when $K=32, M=128$, and $\tau=0.1$. We especially remark the very good agreement between our theoretical analysis and the empirical system performance.

\section{Conclusion}

Conventional RZF precoding provides attractive system throughput in massive MIMO systems, but its computational and implementation complexity is prohibitively high, due to the required channel matrix inversion. In this paper, we have proposed a new class of TPE precoding schemes where the inversion is approximated by truncated polynomial expansions to enable simple hardware implementation. In the single-cell downlink with $M$ transmit antennas and $K$ single-antenna users, this new class can approximate RZF precoding to an arbitrary accuracy by choosing the TPE order $J$ in the interval $1\leq J \leq \min(M,K)$. In terms of implementation complexity, TPE precoding has several advantages: 1) There is no need to compute the precoding matrix beforehand (which leaves more channel uses for data transmission); 2) the delay to the first transmitted symbol is reduced significantly; 3) the multistage structure enables pipelining; and 4) the parameter $J$ can be tailored to the available hardware.

Although the polynomial coefficients depend on the instantaneous channel realizations, we have shown that the per-user SINRs converge to deterministic values in the large-($M,K$) regime. This enabled us to compute asymptotically optimal coefficients using merely the statistics of the channels.
The simulations revealed that the difference in performance between RZF and TPE is small at low SNRs and for large CSI errors. The TPE order $J$ can be chosen very small in these situations and, in general, it does not need to scale with the system dimensions. However, to maintain a fixed per-user rate loss compared to RZF, $J$ should increase with the SNR or as the CSI quality improves.

\appendices

\section{Useful Lemmas}
\label{app:notation}

\begin{lemma}[Common inverses of resolvents] \label{lemma:woodbury}
Given any matrix $\widehat{\bf H} \in \mathbb{C}^{M\times K}$, let $\widehat{\bf h}_k$ denote its $k$th column and $\widehat{\bf H}_k$ denote the matrix obtained after removing the $k$th column from $\widehat{\bf H}$.
The resolvent matrices of $\widehat{\bf H}$ and $\widehat{\bf H}_k$ are denoted by
$	{\bf Q}(t)=\left(\frac{t}{K}\widehat{\bf H}\widehat{\bf H}^{\mbox{\tiny H}}+{\bf I}_M\right)^{-1} $ and $
	{\bf Q}_k(t)=\left(\frac{t}{K}\widehat{\bf H}_k\widehat{\bf H}_k^{\mbox{\tiny H}}+{\bf I}_M\right)^{-1} $
respectively. It then holds, that
\begin{equation}
	{\bf Q}(t)={\bf Q}_k(t)-\frac{1}{K}\frac{t{\bf Q}_k(t)\widehat{\bf h}_k\widehat{\bf h}_k^{\mbox{\tiny H}}{\bf Q}_k(t)}{1+\frac{t}{K}\widehat{\bf h}_k^{\mbox{\tiny H}}{\bf Q}_k(t)\widehat{\bf h}_k}
	\label{eq:Q_k_new}
\end{equation}
and also
\begin{equation}
	{\bf Q}(t)\widehat{\bf h}_k=\frac{{\bf Q}_k(t)\widehat{\bf h}_k}{1+\frac{t}{K}\widehat{\bf h}_k^{\mbox{\tiny H}}{\bf Q}_k(t)\widehat{\bf h}_k} .
	\label{eq:Q_k_h}
\end{equation}
\end{lemma}
\begin{IEEEproof}
This follows from the Woodbury identity \cite{Golub1996a}.
\end{IEEEproof}

The following lemma characterizes the asymptotic behavior of quadratic forms. It will be of frequent use in the computation of deterministic equivalents.
\begin{lemma}[Convergence of quadratic forms]
\label{lemma:convergence_quadratic_form}
Let ${\bf x}_M=\left[X_1,\ldots,X_M\right]^{\mbox{\tiny T}}$ be a $M\times 1$ vector with i.i.d.\ complex Gaussian random variables with unit variance. Let ${\bf A}_M$ be an $M\times M$ matrix independent of ${\bf x}_M$, whose spectral norm is bounded; that is, there exists $C_A < \infty$ such that $\|{\bf A}\|_{2}\leq C_A$. Then, for any $p\geq 1$, there exists a constant $C_p$ depending only on $p$, such that
$$
\mathbb{E}_{{\bf x}_M}\left[\left|\frac{1}{M}{\bf x}_M^{\mbox{\tiny H}}{\bf A}_M{\bf x}_M - \frac{1}{M}\tr ({\bf A}_M) \right|^p \right] \leq \frac{C_pC_A^p}{M^{p/2}}
$$
where the expectation is taken over the distribution of ${\bf x}_M$.
By choosing $p\geq 2$, we thus have that
$$
\frac{1}{M}{\bf x}^{\mbox{\tiny H}}{\bf A}_M{\bf x}-\frac{1}{M}\tr({\bf A}_M) \xrightarrow[M\to+\infty]{\mathrm{a.s.}} 0.
$$
\end{lemma}
\begin{lemma}
Let ${\bf A}_M$ be as in Lemma~\ref{lemma:convergence_quadratic_form}, and ${\bf x}_M,{\bf y}_M$ be random, mutually independent with complex Gaussian entries of zero mean and variance $1$. Then,
$$
\frac{1}{M} {\bf y}_M^{\mbox{\tiny H}}{\bf A}_M{\bf x}_M \xrightarrow[M,K\to+\infty]{\mathrm{a.s.}} 0.
$$
\end{lemma}
\begin{lemma}[Rank-one perturbation lemma]
\label{lemma:perturbation}
Let ${\bf Q}(t)$ and ${\bf Q}_k(t)$ be the resolvent matrices as defined in Lemma~\ref{lemma:woodbury}. Then, for any matrix ${\bf A}$ we have:
$$
\tr \Big( {\bf A}\left({\bf Q}(t)-{\bf Q}_k(t)\right)\Big) \leq \|{\bf A}\|_{2} .
$$
\end{lemma}
\begin{lemma}
	\label{lemma:product}
Let $X_M$ and $Y_M$ be two scalar random variables, with varies such that ${\rm var}(X_M)=\mathcal{O}(M^{-2})$ and ${\rm var}(X_M)=\mathcal{O}(M^{-2})=\mathcal{O}(K^{-2})$. Then
$$
	\mathbb{E}[X_M Y_M]=\mathbb{E}[X_M]\mathbb{E} [Y_M] +o(1).
$$
\end{lemma}
\begin{IEEEproof}
We have
\begin{align*}
	\mathbb{E}[X_M Y_M]&=	\mathbb{E}\left[(X_M -\mathbb{E}[X_M])(Y_M-\mathbb{E}[Y_M])\right]+\mathbb{E[}X_M]\mathbb{E}[Y_M].
\end{align*}
Using the Cauchy-Schwartz inequality, we see that
\begin{align*}
\mathbb{E} \left[ \left|(X_M -\mathbb{E}[X_M])(Y_M-\mathbb{E}[Y_M])\right| \right] & \leq \sqrt{\var(X_M)\var(Y_M)} \\ &=\mathcal{O}(K^{-2})
\end{align*}
which establishes the desired result.
\end{IEEEproof}
\section{Proof of Theorem \ref{th:x_k_z}}
\label{app:x_k_z}
Here we proof Theorem~\ref{th:x_k_z}, which establishes the asymptotic convergence of $X_{k,M}(t,u)$ and $Z_{k,M}(t,u)$ to deterministic quantities.
\subsection{Deterministic equivalent for $X_{k,M}(t,u)$ }
\label{sec:proof_x_km}
 We will begin by treating the random quantity $X_{k,M}(t,u)$. Using the notation of Lemma~\ref{lemma:woodbury}, we can write
$$
X_{k,M}(t,u)=\frac{1}{K^2}{\bf h}_k^{\mbox{\tiny H}}{\bf Q}(t)\widehat{\bf h}_k\widehat{\bf h}_k^{\mbox{\tiny H}}{\bf Q}(u){\bf h}_k .
$$
To control the quadratic form $\frac{1}{K}{\bf h}_k^{\mbox{\tiny H}}{\bf Q}(t)\widehat{\bf h}_k$, we need to remove the dependency of ${\bf Q}(t)$ on vector $\widehat{\bf h}_k$. For that, we shall use the relation in \eqref{eq:Q_k_new}, thereby yielding
\begin{align}
\frac{1}{K}{\bf h}_k^{\mbox{\tiny H}}{\bf Q}(t)\widehat{\bf h}_k&=\frac{1}{K}{\bf h}_k^{\mbox{\tiny H}}{\bf Q}_k(t)\widehat{\bf h}_k \nonumber\\
&-\frac{t}{K^2}\frac{{\bf h}_k^{\mbox{\tiny H}}{\bf Q}_k(t)\widehat{\bf h}_k\widehat{\bf h}_k^{\mbox{\tiny H}}{\bf Q}_k(t)\widehat{\bf h}_k}{1+\frac{t}{K}\widehat{\bf h}_k^{\mbox{\tiny H}}{\bf Q}_k(t)\widehat{\bf h}_k} . \label{eq:decomposition}
\end{align}
Using Lemma~\ref{lemma:convergence_quadratic_form}, we thus have
$$
\frac{1}{K}\widehat{\bf h}_k^{\mbox{\tiny H}}{\bf Q}_k(t)\widehat{\bf h}_k-\frac{1}{K}\tr \left( {\bPhi}{\bf Q}_k(t)  \right) \xrightarrow[M,K\to+\infty]{\mathrm{a.s.}} 0.
$$
Since $\frac{1}{K}\tr \left( \bPhi{\bf Q}_k(t) \right) -\frac{1}{K}\tr \left( \bPhi{\bf Q}(t) \right) \xrightarrow[M,K\to+\infty]{\mathrm{a.s.}} 0$, by the rank-one perturbation property in Lemma~\ref{lemma:perturbation}, we have
 $$
\frac{1}{K}\widehat{\bf h}_k^{\mbox{\tiny H}}{\bf Q}_k(t)\widehat{\bf h}_k-\frac{1}{K} \tr \left( \bPhi{\bf Q}(t) \right) \xrightarrow[M,K\to+\infty]{\mathrm{a.s.}} 0 .
$$
Finally, Theorem~\ref{th:deterministic} implies that
\begin{equation}
\frac{1}{K}\widehat{\bf h}_k^{\mbox{\tiny H}}{\bf Q}_k(t)\widehat{\bf h}_k-\delta(t)\xrightarrow[M,K\to+\infty]{\mathrm{a.s.}} 0.
\label{eq:quadratic_1}
\end{equation}
The same kind of calculations can be used to deal with the quadratic form $\frac{1}{K}{\bf h}_k^{\mbox{\tiny H}}{\bf Q}_k(t)\widehat{\bf h}_k$, whose asymptotic limit is the same as $\frac{\sqrt{1-\tau^2}}{K}\widehat{\bf h}_k^{\mbox{\tiny H}}{\bf Q}_k(t)\widehat{\bf h}_k$, due to the independence between the channel estimation error and the channel vector ${\bf h}_k$. Hence,
\begin{equation}
\frac{1}{K}{\bf h}_k^{\mbox{\tiny H}}{\bf Q}_k(t)\widehat{\bf h}_k-\sqrt{1-\tau^2}\delta(t)\xrightarrow[M,K\to+\infty]{\mathrm{a.s.}} 0.
\label{eq:quadratic_2}
\end{equation}
Plugging the deterministic approximation of  \eqref{eq:quadratic_1} and \eqref{eq:quadratic_2} into \eqref{eq:decomposition}, we thus see that
$$
\frac{1}{K}{\bf h}_k^{\mbox{\tiny H}}{\bf Q}(t)\widehat{\bf h}_k -\frac{\sqrt{1-\tau^2}\delta(t)}{1+t\delta(t)}\xrightarrow[M,K\to+\infty]{\mathrm{a.s.}} 0
$$
and hence
$$
X_{k,M}(t,u)-\frac{(1-\tau^2)\delta(t)\delta(u)}{1+t\delta(t)(1+u\delta(u))}\xrightarrow[M,K\to+\infty]{\mathrm{a.s.}} 0.
$$

\subsection{Deterministic equivalent for $Z_{k,M}(t,u)$ }
Finding a deterministic equivalent for $Z_{k,M}(t,u)$ is much more involved than for $X_{k,M}(t,u)$. Following the same steps as in Appendix~\ref{sec:proof_x_km}, we decompose $Z_{k,M}(t,u)$ as
\begin{align*}
	&Z_{k,M}(t,u)=\frac{1}{K}{\bf h}_k^{\mbox{\tiny H}} {\bf Q}_k(t){\widehat{\bf H}{\bf P}\widehat{\bf H}^{\mbox{\tiny H}}}{\bf Q}_k(u){\bf h}_k\\
 &-\frac{\frac{u}{K^2}{\bf h}_k^{\mbox{\tiny H}}{\bf Q}_k(t)\widehat{\bf H}{\bf P}\widehat{\bf H}^{\mbox{\tiny H}}{\bf Q}_k(u)\widehat{\bf h}_k\widehat{\bf h}_k^{\mbox{\tiny H}}{\bf Q}_k(u){\bf h}_k}{1+\frac{u}{K}\widehat{\bf h}_k^{\mbox{\tiny H}}{\bf Q}_k(u)\widehat{\bf h}_k} \\
&-\frac{\frac{t}{K^2}{\bf h}_k^{\mbox{\tiny H}}{\bf Q}_k(t)\widehat{\bf h}_k\widehat{\bf h}_k^{\mbox{\tiny H}}{\bf Q}_k(t)\widehat{\bf H}{\bf P}\widehat{\bf H}^{\mbox{\tiny H}}{\bf Q}_k(u){\bf h}_k}{1+\frac{t}{K}\widehat{\bf h}_k^{\mbox{\tiny H}}{\bf Q}_k(t)\widehat{\bf h}_k}\\
&+\frac{\frac{tu}{K^3}{\bf h}_k^{\mbox{\tiny H}}{\bf Q}_k(t)\widehat{\bf h}_k\widehat{\bf h}_k^{\mbox{\tiny H}}{\bf Q}_k(t)\widehat{\bf H}{\bf P}\widehat{\bf H}^{\mbox{\tiny H}}{\bf Q}_k(u)\widehat{\bf h}_k\widehat{\bf h}_k^{\mbox{\tiny H}}{\bf Q}_k(u){\bf h}_k}{(1+\frac{t}{K}\widehat{\bf h}_k^{\mbox{\tiny H}}{\bf Q}_k(t)\widehat{\bf h}_k)(1+\frac{u}{K}\widehat{\bf h}_k^{\mbox{\tiny H}}{\bf Q}_k(u)\widehat{\bf h}_k)}\\
       &\triangleq X_1(t,u)+X_2(t,u)+X_3(t,u)+X_4(t,u).
\end{align*}
As it will be shown next, to determine the asymptotic limit of the random variables $X_i(t,u),i=1,\ldots,4$, we need to find a deterministic equivalent for
$$
\frac{1}{K}\tr\left(\bPhi{\bf Q}(t)\widehat{\bf H}{\bf P}\widehat{\bf H}^{\mbox{\tiny H}}{\bf Q}(u)\right).
$$
This is the most involved step of the proof. It will, thus, be treated separately in Appendix~\ref{app:alpha}, where we establish the following lemma:
\begin{lemma}
\label{lem:alphabar} \label{lemma:new_deter}
	Let ${\bf H}$ be an $M\times K$ random matrix whose columns are drawn according to Assumption A-\ref{ass:channel}. Define for $t\geq 0$, the resolvent matrix
	$
	{\bf Q}(t)=\left(\frac{t}{K}{\bf H}{\bf H}^{\mbox{\tiny H}}+{\bf I}_K\right)^{-1} .
	$
	Let ${\bf A}$ be an $M\times M$ deterministic matrix with uniformly spectral norm and $\widehat{\alpha}_M(t,u,{\bf A})$  given as
	$$
	\widehat{\alpha}_{M}(t,u,{\bf A})=\frac{1}{K}\tr\left({\bf A}{\bf Q}(t){\bf H}{\bf P}{\bf H}^{\mbox{\tiny H}}{\bf Q}(u)\right) .
	$$
	Then, in the asymptotic regime described by Assumption A-\ref{ass:regime}, we have
	$$
	\widehat{\alpha}_M(t,u,{\bf A})-\overline{\alpha}_M(t,u,{\bf A})\xrightarrow[M,K\to+\infty]{\mathrm{a.s.}}0
	$$
where 
\begin{align}
	& \overline{\alpha}_M(t,u,{\bf A})=\tr ({\bf P}) \frac{\frac{1}{K} \tr \left( \bPhi{\bf T}(u){\bf A}{\bf T}(t) \right) }{(1+t\delta(t))(1+u\delta(u))} \nonumber\\
											  &\quad +\frac{\tr({\bf P})}{(1+t\delta(t))(1+u\delta(u))} \nonumber\\
				 &\quad \times \frac{\frac{tu}{K}\tr \left( \bPhi{\bf T}(u){\bf A}{\bf T}(t) \right) \frac{1}{K}\tr \left( \bPhi{\bf T}(u)\bPhi{\bf T}(t) \right)}{(1+t\delta(t))(1+u\delta(u))-\frac{tu}{K}\tr \left(\bPhi{\bf T}(u)\bPhi{\bf T}(t)\right)}\label{eq:result_alpha_A} .
\end{align}
In particular, if ${\bf A}=\bPhi$, we have
$$
\overline{\alpha}_M(t,u,\bPhi)=\frac{\tr({\bf P})\frac{1}{K}\tr \left( \bPhi{\bf T}(u)\bPhi{\bf T}(t) \right) }{(1+t\delta(t))(1+u\delta(u))-\frac{tu}{K}\tr \left( \bPhi{\bf T}(u)\bPhi{\bf T}(t) \right)} .
$$
The proof of this lemma is adjourned to Appendix~\ref{app:alpha}.
\end{lemma}

Let us begin by treating $X_1(t,u)$:
\begin{align*}
\frac{1}{K}{\bf h}_k^{\mbox{\tiny H}}{\bf Q}_k(t)\widehat{\bf H}{\bf P}\widehat{\bf H}^{\mbox{\tiny H}}{\bf Q}_k(u){\bf h}_k&=\frac{1}{K}{\bf h}_k^{\mbox{\tiny h}}{\bf Q}_k(t){\widehat{\bf H}_k{\bf P}_k\widehat{\bf H}_k}{\bf Q}_k(u){\bf h_k} \\		     &+\frac{p_k}{K}{\bf h}_k^{\mbox{\tiny H}}{\bf Q}_k(t)\widehat{\bf h}_k\widehat{\bf h}_k^{\mbox{\tiny H}}{\bf Q}_k(u){\bf h}_k .
\end{align*}
The right-hand side term in the equation above can be treated using \eqref{eq:quadratic_2}, thereby yielding
$$
\frac{p_k}{K}{\bf h}_k^{\mbox{\tiny H}}{\bf Q}_k(t)\widehat{\bf h}_k\widehat{\bf h}_k^{\mbox{\tiny H}}{\bf Q}_k(u){\bf h}_k-Kp_k(1-\tau^2)\delta(t)\delta(u)\xrightarrow[M,K\to\infty]{\mathrm{a.s.}} 0.
$$
Using Lemma~\ref{lemma:convergence_quadratic_form}, we can prove that
\begin{align}
	&\frac{1}{K}{\bf h}_k^{\mbox{\tiny H}}{\bf Q}_k(t)\widehat{\bf H}_k{\bf P}_k\widehat{\bf H}_k^{\mbox{\tiny H}}{\bf Q}_k(u){\bf h_k}\nonumber\\
	&-\frac{1}{K}\tr\left(\bPhi{\bf Q}_k(t){\widehat{\bf H}_k{\bf P}_k\widehat{\bf H}_k^{\mbox{\tiny H}}}{\bf Q}_k(u)\right)
	\xrightarrow[M,K\to+\infty]{\mathrm{a.s.}} 0 . \label{eq:conv_1}
\end{align}
Continuing, according to Lemma~\ref{lemma:new_deter}, we have
\begin{align}
	&\frac{1}{K}\tr\left(\bPhi{\bf Q}_k(t){\widehat{\bf H}_k{\bf P}_k\widehat{\bf H}_k^{\mbox{\tiny H}}}{\bf Q}_k(u)\right)-\tr \left({\bf P}\right)\beta_{M}(t,u) \nonumber \\ &\qquad \xrightarrow[M,K\to+\infty]{\mathrm{a.s.}}0.\label{eq:new_deter}
\end{align}
Combining \eqref{eq:conv_1} with \eqref{eq:new_deter} yields
\begin{align*}
&\frac{1}{K}{\bf h}_k^{\mbox{\tiny H}}{\bf Q}_k(t)\widehat{\bf H}_k{\bf P}_k\widehat{\bf H}_k^{\mbox{\tiny H}}{\bf Q}_k(u){\bf h}_k-\tr\left({\bf P}\right)\beta_M(t,u) \xrightarrow[M,K\to+\infty]{\mathrm{a.s.}}0.
\end{align*}
Thus, in the asymptotic regime we have 
\begin{align}
& X_1(t,u)-\left(Kp_k(1-\tau^2)\delta(t)\delta(u)+\tr ({\bf P}) \beta_M(t,u)\right) \nonumber \\ & \qquad \xrightarrow[M,K\to+\infty]{\mathrm{a.s.}}0.
\label{eq:x_1}
\end{align}
Controlling the other terms $X_i(t,u),i=2,3,4$, will also include the term $\beta(t,u)$. First note that $X_2(t,u)$ is given by
\begin{align*}
	X_{2}(t,u)=-uY_2(t,u)\frac{\frac{1}{K}\widehat{\bf h}_k^{\mbox{\tiny H}}{\bf Q}_k(u){\bf h}_k}{1+\frac{u}{K}\widehat{\bf h}_k^{\mbox{\tiny H}}{\bf Q}_k{\widehat{\bf h}_k}}
\end{align*}

where
$$
Y_2(t,u)=\frac{1}{K}{\bf h}_k^{\mbox{\tiny H}}{\bf Q}_k(t)\widehat{\bf H}{\bf P}\widehat{\bf H}^{\mbox{\tiny H}}{\bf Q}_k(u)\widehat{\bf h}_k.
$$
Observe that $Y_2(t,u)$  is very similar to $X_1(t,u)$. The only difference is that $Y_2(t,u)$ is a quadratic form involving vectors ${\bf h}_k$ and $\widehat{\bf h}_k$ whereas $X_1(t,u)$ involves only the vector ${\bf h}_k$. Following the same kind of calculations leads to
\begin{align*}
	&Y_2(t,u)-\left(Kp_k\sqrt{1-\tau^2}\delta(t)\delta(u)+\sqrt{1-\tau^2}\tr \left( {\bf P}\right)\beta_M(t,u)\right)\\
	&\qquad \xrightarrow[M,K\to+\infty]{\mathrm{a.s.}}0.
\end{align*}
Since $\frac{\frac{1}{K}\widehat{\bf h}_k^{\mbox{\tiny H}}{\bf Q}_k(u){\bf h}_k}{1+\frac{u}{K}\widehat{\bf h}_k{\bf Q}_k(u)\widehat{\bf h}_k}$ satisfies
$$
\frac{\frac{1}{K}\widehat{\bf h}_k^{\mbox{\tiny H}}{\bf Q}_k(u){\bf h}_k}{1+\frac{u}{K}\widehat{\bf h}_k{\bf Q}_k(u)\widehat{\bf h}_k}-\frac{\sqrt{1-\tau^2}\delta(u)}{1+u\delta(u)} \xrightarrow[M,K\to+\infty]{\mathrm{a.s.}}0
$$
we now have
\begin{small}
\begin{align}
	&X_2(t,u)+\frac{u\delta(u)\left(Kp_k(1-\tau^2)\delta(t)\delta(u)+({1-\tau^2})\tr \left({\bf P}\right)\beta_M(t,u)\right)}{1+u\delta(u)}\nonumber\\
	&\qquad\xrightarrow[M,K\to+\infty]{\mathrm{a.s.}}0.\label{eq:x_2}
\end{align}
\end{small}
Similarly, $X_3(t,u)$ satisfies
\begin{small}
\begin{align}
	&X_3(t,u)+\frac{t\delta(t)\left(Kp_k(1-\tau^2)\delta(t)\delta(u)+({1-\tau^2})\tr \left({\bf P}\right)\beta_M(t,u)\right)}{1+t\delta(t)}\nonumber\\
	&\qquad\xrightarrow[M,K\to+\infty]{\mathrm{a.s.}}0.\label{eq:x_3}
\end{align}
\end{small}

Finally, $X_4(t,u)$ can be treated using the same approach, thereby providing the following convergence:
\begin{align}
	&X_4(t,u)-\frac{tu\delta(t)\delta(u)(1-\tau^2)\left(Kp_k\delta(t)\delta(u)+\tr \left({\bf P}\right)\beta_M(t,u)\right)}{(1+t\delta(t))(1+u\delta(u))} \nonumber\\ & \qquad\xrightarrow[M,K\to+\infty]{\mathrm{a.s.}}0.\label{eq:x_4}
\end{align}
Summing \eqref{eq:x_1}, \eqref{eq:x_2}, \eqref{eq:x_3}, \eqref{eq:x_4} yields
\begin{align*}
	&Z_{k,M}(t,u)-\left(\frac{K p_k(1-\tau^2)\delta(t)\delta(u)}{(1+t\delta(t))(1+u\delta(u))}\right.\\
&\left.+\tr \left({\bf P}\right) \left(\tau^2+\frac{(1-\tau^2)}{(1+u\delta(u))(1+t\delta(t))}\right)\beta_M(t,u)\right) \xrightarrow[M,K\to+\infty]{\mathrm{a.s.}}0.
\end{align*}

\section{Proof of Lemma~\ref{lemma:new_deter}}
\label{app:alpha}
The aim of this section is to determine a deterministic equivalent for the random quantity
$$
\widehat{\alpha}_M(t,u,{\bf A})=\frac{1}{K}\tr \left( {\bf A}{\bf Q}(t){\bf H}{\bf P}{\bf H}^{\mbox{\tiny H}}{\bf Q}(u) \right).
$$
The proof is technical and will make frequent use of results from Appendix~\ref{app:notation}. First, we need to control $\var\left({\widehat{\alpha}}_M(t,u)\right)$. This has already been treated in \cite{HAC06} where it was proved that $\var\left(\widehat{\alpha}_M(t,u,{\bf A})\right)=\mathcal{O}(K^{-2})$ when $t=u$. The same calculations hold for $t\neq u$, thus we consider in the sequel that ${\rm var}\left(\widehat{\alpha}_M(t,u,{\bf A})\right)=\mathcal{O}(K^{-2}).$
Hence, we have
\begin{equation}
	\widehat{\alpha}_M(t,u,{\bf A})-\mathbb{E}[\widehat{\alpha}_M(t,u,{\bf A})]\xrightarrow[M,K\to+\infty]{\mathrm{a.s.}}0.
\label{eq:derivation}
\end{equation}
Equation~\eqref{eq:derivation} allows us to focus directly on controlling $\mathbb{E}[\widehat{\alpha}_M(t,u,{\bf A})]$.
Using the resolvent identity
\begin{align*}
	{\bf Q}(t)-{\bf T}(t)&={\bf T}(t)\left({\bf T}^{-1}(t)-{\bf Q}^{-1}(t)\right){\bf Q}(t)\\
	&={\bf T}(t)\left(\frac{t\bPhi}{1+t\delta(t)}-\frac{t}{K}{\bf H}{\bf H}^{\mbox{\tiny H}}\right){\bf Q}(t)
\end{align*}
we decompose $\widehat{\alpha}_M(t,u,{\bf A})$ as
\begin{align*}
	\widehat{\alpha}_M(t,u,{\bf A})&=\frac{1}{K}\tr \big( {\bf A}{\bf T}(t){\bf H}{\bf P}{\bf H}^{\mbox{\tiny H}}{\bf Q}(u) \big) \\
	 &+\frac{t\tr \big( {\bf A}{\bf T}(t)\bPhi{\bf Q}(t){\bf H}{\bf P}{\bf H}^{\mbox{\tiny H}}{\bf Q}(u) \big)}{K(1+t\delta(t))} \\
	 &-\frac{t}{K^2}\tr \big( {\bf A}{\bf T}(t){\bf H}{\bf H}^{\mbox{\tiny H}}{\bf Q}(t){\bf H}{\bf P}{\bf H}^{\mbox{\tiny H}}{\bf Q}(u) \big) \\
	&=Z_1+Z_2+Z_3.
\end{align*}
We will only directly deal with the terms $Z_1$ and $Z_3$, since $Z_2$ will be compensated by terms in $Z_3$. We begin with $Z_1$:
\begin{align*}
&\mathbb{E}\left[Z_1\right]=\frac{1}{K} \sum_{\ell=1}^K p_\ell\mathbb{E}\left[\tr \big( {\bf A}{\bf T}(t){\bf h}_\ell{\bf h}_\ell^{\mbox{\tiny H}}{\bf Q}(u) \big) \right]\\
								      &=\frac{1}{K}\sum_{\ell=1}^K p_\ell\mathbb{E}\left[\frac{{\bf h}_\ell^{\mbox{\tiny H}}{\bf Q}_\ell(u){\bf A}{\bf T}(t){\bf h}_\ell}{1+\frac{u}{K}{\bf h}_\ell^{\mbox{\tiny H}}{\bf Q}_\ell(u){\bf h}_\ell}\right]\\
							       &=\sum_{\ell=1}^K \frac{p_\ell}{K}\mathbb{E}\left[\frac{{\bf h}_\ell^{\mbox{\tiny H}}{\bf Q}_\ell(u){\bf A}{\bf T}(t){\bf h}_\ell\left(\frac{u}{K}\tr \big( \bPhi{\bf Q}_\ell \big) -\frac{u}{K}{\bf h}_\ell^{\mbox{\tiny H}}{\bf Q}_\ell(u){\bf h}_\ell\right)}{\left(1+\frac{u}{K}{\bf h}_\ell^{\mbox{\tiny H}}{\bf Q}_\ell(u){\bf h}_\ell\right)\left(1+\frac{u}{K}\tr\bPhi{\bf Q}_\ell(u)\right)}\right]\\
					 &\qquad +\frac{p_\ell}{K}\mathbb{E}\left[\frac{{\bf h}_\ell^{\mbox{\tiny H}}{\bf Q}_\ell(u){\bf A}{\bf T}(t){\bf h}_\ell}{1+\frac{u}{K}\tr\bPhi{\bf Q}_\ell(u)}\right] .
\end{align*}
Using Lemma~\ref{lemma:convergence_quadratic_form}, we can show that the first term on the right hand side of the above equation is negligible. Therefore,
\begin{align*}
	\mathbb{E}\left[Z_1\right]&=\sum_{\ell=1}^K \frac{p_\ell}{K}\mathbb{E}\left[\frac{{\bf h}_\ell^{\mbox{\tiny H}}{\bf Q}_\ell(u){\bf A}{\bf T}(t){\bf h}_\ell}{1+\frac{u}{K}\tr \big( \bPhi{\bf Q}_\ell(u) \big) } \right]+{o}(1)\\
	&=\sum_{\ell=1}^K \frac{p_\ell}{K}\mathbb{E}\left[\frac{\tr\bPhi{\bf Q}_\ell(u){\bf A}{\bf T}(t)}{1+\frac{u}{K}\tr \big( \bPhi{\bf Q}_\ell \big)}\right]+{o}(1) .
\end{align*}
Using Lemma~\ref{lemma:perturbation}, we have
$$
E\left[Z_1\right]=\sum_{\ell=1}^K \frac{p_\ell}{K}\mathbb{E}\left[\frac{\tr \big( \bPhi{\bf Q}(u){\bf A}{\bf T}(t) \big)}{1+\frac{u}{K}\tr \big( \bPhi{\bf Q}(u) \big)}\right]+{o}(1) .
$$
Theorem~\ref{th:deterministic}, thus, implies
\begin{align*}
	E\left[Z_1\right]&=\sum_{\ell=1}^K \frac{p_\ell}{K}\mathbb{E}\left[\frac{\tr \big( \bPhi{\bf T}(u){\bf A}{\bf T}(t)\big)}{\big(1+u\delta(u) \big) }\right]+{o}(1)\\
											       &=\frac{\frac{1}{K}\tr({\bf P}) \frac{1}{K}\tr \big( \bPhi{\bf T}(u){\bf A}{\bf T}(t) \big) }{1+u\delta(u)}+o(1).
\end{align*}
We now look at $Z_3$, where 
\begin{align*}
	Z_3&=-\frac{t}{K^2}\sum_{\ell=1}^K \tr\left({\bf A}{\bf T}(t){\bf h}_\ell{\bf h}_\ell^{\mbox{\tiny H}}{\bf Q}(t){\bf H}{\bf P}{\bf H}^{\mbox{\tiny H}}{\bf Q}(u)\right) .
\end{align*}
Using \eqref{eq:Q_k_h}, we arrive at
$$
Z_3=-\frac{t}{K^2}\sum_{\ell=1}^K \frac{ \tr\big( {\bf A}{\bf T}(t){\bf h}_\ell{\bf h}_\ell^{\mbox{\tiny H}}{\bf Q}_\ell(t){\bf H}{\bf P}{\bf H}^{\mbox{\tiny H}}{\bf Q}(u) \big)}{1+\frac{t}{K}{\bf h}_\ell^{\mbox{\tiny H}}{\bf Q}_\ell(t){\bf h}_\ell} .
$$
From \eqref{eq:Q_k_new}, $Z_3$ can be decomposed as
\begin{align*}
	Z_3&=-\frac{t}{K^2}\sum_{\ell=1}^K \frac{ \tr\left({\bf A}{\bf T}(t){\bf h}_\ell{\bf h}_\ell^{\mbox{\tiny H}}{\bf Q}_\ell(t){\bf H}{\bf P}{\bf H}^{\mbox{\tiny H}}{\bf Q}_\ell(u) \right)}{1+\frac{t}{K}{\bf h}_\ell^{\mbox{\tiny H}}{\bf Q}_\ell(t){\bf h}_\ell}\\
	   &+\frac{tu}{K^3}\sum_{\ell=1}^K \frac{ \tr\left({\bf A}{\bf T}(t){\bf h}_\ell{\bf h}_\ell^{\mbox{\tiny H}}{\bf Q}_\ell(t){\bf H}{\bf P}{\bf H}^{\mbox{\tiny H}}{\bf Q}_\ell(u){\bf h}_\ell{\bf h}_\ell^{\mbox{\tiny H}}{\bf Q}_\ell(u) \right)}{\left(1+\frac{t}{K}{\bf h}_\ell^{\mbox{\tiny H}}{\bf Q}_\ell(t){\bf h}_\ell\right)\left(1+\frac{u}{K}{\bf h}_\ell^{\mbox{\tiny H}}{\bf Q}_\ell(u){\bf h}_\ell\right)}\\
	   &=Z_{31}+Z_{32} .
\end{align*}
We sequentially deal with the terms $Z_{31}$ and $Z_{32}$.
The same arguments as those used before, allow us to substitute the denominator by $1+t\delta(t)$, thereby yielding:
\begin{align*}
	&\mathbb{E}\left[Z_{31}\right]=-\frac{t}{K^2}\sum_{\ell=1}^K\mathbb{E}\left[ \frac{{\bf h}_\ell^{\mbox{\tiny H}}{\bf Q}_\ell(t){\bf H}{\bf P}{\bf H}^{\mbox{\tiny H}}{\bf Q}_\ell(u){\bf A}{\bf T}(t){\bf h}_\ell}{1+t\delta(t)}\right]+o(1)\\
 &=-\frac{t}{K^2}\left(\sum_{\ell=1}^K  \mathbb{E}\left[\frac{{\bf h}_\ell^{\mbox{\tiny H}}{\bf Q}_\ell(t){\bf H}_\ell{\bf P}_\ell{\bf H}_\ell^{\mbox{\tiny H}}{\bf Q}_\ell(u){\bf A}{\bf T}(t){\bf h}_\ell}{1+t\delta(t)}\right]\right. \\
 &\qquad \left.+p_\ell\mathbb{E}\left[\frac{{\bf h}_\ell^{\mbox{\tiny H}}{\bf Q}_\ell(t){\bf h}_\ell{\bf h}_\ell^{\mbox{\tiny H}}{\bf Q}_\ell(u){\bf A}{\bf T}(t){\bf h}_\ell}{1+t\delta(t)}\right]\right)+o(1)\\
 &=-\frac{t}{K^2}\left(\sum_{\ell=1}^K\mathbb{E}\left[\frac{\tr \big( \bPhi{\bf Q}_\ell(t){\bf H}_\ell{\bf P}_\ell{\bf H}_\ell^{\mbox{\tiny H}}{\bf Q}_\ell(u){\bf A}{\bf T}(t) \big) }{1+t\delta(t)}\right]\right.\\
 &\qquad\left.+p_\ell\mathbb{E}\left[\frac{{\bf h}_\ell^{\mbox{\tiny H}}{\bf Q}_\ell(t){\bf h}_\ell{\bf h}_\ell^{\mbox{\tiny H}}{\bf Q}_\ell(u){\bf A}{\bf T}(t){\bf h}_\ell }{1+t\delta(t)}\right]\right)+o(1)\\
	&\triangleq\chi_1+\chi_2 .
\end{align*}
By Lemma~\ref{lemma:convergence_quadratic_form}, the quadratic forms involved in $\chi_2$ have variance $\mathcal{O}(K^{-2})$, and thus can be substituted by their expected mean (see Lemma~\ref{lemma:product}). We obtain
\begin{align}
	\chi_2&=-t\sum_{\ell=1}^K p_\ell\mathbb{E}\left[\frac{\frac{1}{K}\tr \big( \bPhi{\bf Q}_\ell(t) \big) \frac{1}{K}\tr \big(\bPhi{\bf Q}_\ell(u){\bf A}{\bf T}(t) \big)}{1+t\delta(t)}\right] +o(1)\nonumber\\
	      &=-\frac{t\delta(t)}{1+t\delta(t)}\tr({\bf P}) \frac{1}{K}\tr \big( \bPhi{\bf T}(u){\bf A}{\bf T}(t)\big)+o(1) \label{eq:chi_2}.	
\end{align}
The term $\chi_1$ will be compensated by ${Z_2}$. To see that, observe that the first order of $\chi_1$ does not change if we substitute ${\bf H}_\ell$ by ${\bf H}$ and ${\bf P}_\ell$ by ${\bf P}$.  Besides, due to Lemma~\ref{lemma:perturbation}, we can  substitute ${\bf Q}_\ell(t)$ by ${\bf Q}(t)$ and ${\bf Q}_\ell(u)$ by ${\bf Q}(u)$, hence proving that
\begin{equation}
\chi_1=-\mathbb{E}\left[Z_2\right]+o(1).
\label{eq:chi_1}
\end{equation}
Finally, it remains to deal with $Z_{32}$. Substituting $\frac{1}{K}{\bf h}_\ell^{\mbox{\tiny H}}{\bf Q}_\ell(t){\bf h}_\ell$ and $\frac{1}{K}{\bf h}_\ell^{\mbox{\tiny H}}{\bf Q}_\ell(u){\bf h}_\ell$ by their asymptotic equivalent $\delta(t)$ and $\delta(u)$, we get
\begin{align*}
	& \mathbb{E}\left[Z_{32}\right]=\\
	& \frac{tu}{K^3}\sum_{\ell=1}^K \mathbb{E}\left[\frac{{\bf h}_\ell^{\mbox{\tiny H}}{\bf Q}_\ell(u){\bf A}{\bf T}(t){\bf h}_\ell {\bf h}_\ell^{\mbox{\tiny H}}{\bf Q}_\ell(t){\bf H}_\ell{\bf P}_\ell{\bf H}_\ell^{\mbox{\tiny H}}{\bf Q}_\ell(u){\bf h}_\ell}{(1+t\delta(t))(1+u\delta(u))}\right] + \\
 	&\frac{tu}{K^3}\sum_{\ell=1}^K p_\ell\mathbb{E}\left[\frac{{\bf h}_\ell^{\mbox{\tiny H}}{\bf Q}_\ell(u){\bf A}{\bf T}(t){\bf h}_\ell{\bf h}_\ell^{\mbox{\tiny H}}{\bf Q}_\ell(t){\bf h}_\ell{\bf h}_\ell^{\mbox{\tiny H}}{\bf Q}_\ell(u){\bf h}_\ell}{(1+t\delta(t))(1+u\delta(u))}\right] +o(1).
\end{align*}
Analogously to before, $\mathbb{E}\left[Z_{32}\right]$ can be simplified: 
\begin{align}
	&\mathbb{E}\left[Z_{32}\right]=\frac{tu}{K^3}\sum_{\ell=1}^K \mathbb{E}\left[\frac{\tr \big( \bPhi{\bf Q}(t){\bf H}{\bf P}{\bf H}^{\mbox{\tiny H}}{\bf Q}(u) \big) \tr \big( \bPhi{\bf T}(u){\bf A}{\bf T}(t) \big)}{(1+t\delta(t))(1+u\delta(u))}\right] \nonumber\\
 &\qquad+\frac{tu}{K}\sum_{\ell=1}^K \frac{p_\ell \delta(t)\delta(u) \tr \big(\bPhi{\bf T}(u){\bf A}{\bf T}(t) \big)}{(1+t\delta(t))(1+u\delta(u))}+o(1)\nonumber\\
	&=\frac{tu}{K}\frac{\tr \big( \bPhi{\bf T}(u){\bf A}{\bf T}(t) \big) \mathbb{E} [\widehat{\alpha}_{M}(t,u,\bPhi) ]}{(1+t\delta(t))(1+u\delta(u))} \nonumber\\
 &	\qquad+\frac{\delta(t)\delta(u)\tr ( {\bf P}) \frac{tu}{K}\tr \big( \bPhi{\bf T}(u){\bf A}{\bf T}(t) \big)}{(1+t\delta(t))(1+u\delta(u))} +o(1)\label{eq:Z_32}.
\end{align}
Combining \eqref{eq:chi_2}, \eqref{eq:chi_1} and \eqref{eq:Z_32}, we obtain
\begin{align}
	&\mathbb{E}[\widehat{\alpha}_M(t,u,{\bf A})]=\frac{\tr ({\bf P}) \frac{1}{K}\tr \big( \bPhi{\bf T}(u){\bf A}{\bf T}(t) \big) }{(1+t\delta(t))(1+u\delta(u))}\nonumber\\
	&\qquad+\frac{tu}{K}\frac{\tr \big( \bPhi{\bf T}(u){\bf A}{\bf T}(t) \big) \mathbb{E} [\widehat{\alpha}_{M}(t,u,\bPhi)] }{(1+t\delta(t))(1+u\delta(u))}+o(1). \label{eq:alpha_A}
\end{align}
Replacing ${\bf A}$ with $\bPhi$, one finds a deterministic equivalent
\begin{align}
	&\mathbb{E}[\widehat{\alpha}_M(t,u,\boldsymbol{\Phi})]= \nonumber\\
	&\frac{\tr({\bf P}) \frac{1}{K}\tr \big(\bPhi{\bf T}(u)\bPhi{\bf T}(t) \big) }{(1+t\delta(t))(1+u\delta(u))-\frac{tu}{K}\tr \big(\bPhi{\bf T}(u)\bPhi{\bf T}(t) \big)}
	+o(1) . \label{eq:alpha_phi}
\end{align}
Finally, substituting \eqref{eq:alpha_phi} into \eqref{eq:alpha_A} establishes \eqref{eq:result_alpha_A}.

\section{Proof of Corollary~\ref{corollary:derivative}}
\label{app:derivative}
The proof of Corollary~\ref{corollary:derivative} relies on Montel's theorem \cite{RUD86}. We  only prove the result for $X_{k,M}(t,u)$, $Z_{k,M}(t,u)$ follows analogously. Note, that $X_{k,M}(t,u)$ and $\overline{X}_{k,M}(t,u)$ are analytic functions, when their domains are extended to $\mathbb{C}\backslash\mathbb{R}_{-}\times \mathbb{C}\backslash\mathbb{R}_{-}$, where $\mathbb{R}_{-}$ is the set of negative real-valued numbers.
Since $X_{k,M}(t,u)-\overline{X}_{k,M}(t,u)$ is almost surely bounded for large $M$ and $K$ on every compact subset of $\mathbb{C}\backslash\mathbb{R}_{-}$, Montel's theorem asserts that there exists a converging subsequence, which converges to an analytic function. Since this limiting function is necessarily zero on the positive real axis, it must be zero everywhere. 
Thus, from every subsequence one can extract a convergent one that converges to zero, thus
\begin{equation}
X_{k,M}(z_1,z_2)-\overline{X}_{k,M}(z_1,z_2)\xrightarrow[M,K\to+\infty]{\mathrm{a.s.}}0 \   \  \forall   z_1, z_2 \in \mathbb{C}\backslash\mathbb{R}_{-}
\label{eq:x_k_m_d}
\end{equation}
As $X_{k,M}(z_1,z_2)$ is analytic, the derivatives of $X_{k,M}(z_1,z_2)-\overline{X}_{k,M}(z_1,z_2)$ converge to zero. In particular, if $\tilde{t}$ and $\tilde{u}$ are strictly positive scalars, we have
\begin{equation}
X_{k,M}^{(m,\ell)}(\tilde{t},\tilde{u}) - \overline{X}_{k,M}^{(m,\ell)}(\tilde{t},\tilde{u}) \xrightarrow[M,K\to+\infty]{\mathrm{a.s.}}0.
\label{eq:convergence_zero}
\end{equation}
This result can be extended to the case of $\tilde{t}=0$ and $\tilde{u}=0$. To see this, let $\eta>0$ and decompose
\begin{align*}
	X_{k,M}^{(m,\ell)} - \overline{X}_{k,M}^{(m,\ell)} = \alpha_1+\alpha_2+\alpha_3
\end{align*}
where
\begin{align*}
	\alpha_1&=X_{k,M}^{(m,\ell)} - X_{k,M}^{(m,\ell)}(\eta,\eta)\\
	\alpha_2&=X_{k,M}^{(m,\ell)}(\eta,\eta) - \overline{X}_{k,M}^{(m,\ell)}(\eta,\eta)\\
	\alpha_3&=\overline{X}_{k,M}^{(m,\ell)}(\eta,\eta) - \overline{X}_{k,M}^{(m,\ell)} .
\end{align*}
Now, let $\epsilon>0$. Since the derivatives of $X_{k,M}^{(m,\ell)}$ and $\overline{X}_{k,M}^{(m,\ell)}$ are almost surely bounded for large $M$ and $K$, the quantities $|\alpha_1|$ and $|\alpha_3|$ can be made smaller than $\epsilon/3$ when $\eta$ is small enough. On the other hand, \eqref{eq:convergence_zero} implies that $\alpha_2$ converges to zero almost surely. There exists $M_0$, such that, for $M\geq M_0$ we have
$|\alpha_2|\leq \frac{\epsilon}{3}$.
Therefore, for $M$ large enough,
$
\left| X_{k,M}^{(m,\ell)} - \overline{X}_{k,M}^{(m,\ell)} \right|\leq \epsilon, 
$
thereby proving
$$
X_{k,M}^{(m,\ell)} - \overline{X}_{k,M}^{(m,\ell)} \xrightarrow[M,K\to+\infty]{\mathrm{a.s.}}0.
$$

\section{Iterative Algorithm for Computing $\beta_M^{(\ell,m)}$}
\label{app:alg_iterbeta}
An iterative approach for computing $\beta_M^{(\ell,m)}$ is given by 
\begin{algorithm}[H]
\caption{Iterative algorithm for the computation of $\beta_M^{(\ell,m)}$}
\begin{algorithmic}
\For {$k=0 \to J$}
\State $\beta_M^{(k,0)}\gets \frac{1}{K}\tr \big(\bPhi\Tau^{(k)}\bPhi \big)$, $\beta_M^{(0,k)}\gets \frac{1}{K}\tr \big(\bPhi\Tau^{(k)}\bPhi \big)$
\EndFor
\For {$m=1\to J$}
\For {$k=1\to J$}
\State $\beta_M^{(k,m)}\gets \frac{1}{K}\tr \big( \bPhi\Tau^{(k)}\bPhi\Tau^{(m)}\bPhi \big)$
\For {$p_k=1\to k$}
\For {$q_m=1\to m$}
\State $\beta_M^{(k,m)} \gets \beta_M^{(k,m)} - $ \\ \hspace{0.2cm} $p_k q_m {k\choose p_k}{m \choose q_m}\beta_M^{(p_k-1,q_m-1)}\frac{1}{K}\tr \big( \bPhi\Tau^{(k-p_k)}\bPhi\Tau^{(m-q_m)} \big)$
\EndFor
\EndFor
\EndFor
\EndFor
\end{algorithmic}
\end{algorithm}

\section{Iterative Algorithm for Computing ${\bf T}^{(q)}$}
\label{app:IterAlgoT}
For the sake of completeness, we provide hereafter an algorithm that can be used to compute ${\bf T}^{(q)}$. It is an adapted version of the iterative algorithm given in \cite{hoydis}.
\begin{algorithm}[H]
\begin{algorithmic}
	\State $\delta^{(0)}\gets \frac{1}{K}\tr(\boldsymbol{\Phi})$
	\State $g^{(0)} \gets 0$
	\State $f^{(0)} \gets -\frac{1}{1+g^{(0)}}$
	\State ${\bf T}^{(0)} \gets {\bf I}_M$
	\State ${\bf R}^{(0)} \gets {\bf 0}_M$
	\For {$i=1 \to p$}
	\State ${\bf R}^{(i)}\gets if^{(i-1)}\boldsymbol{\Phi}$
	\State ${\bf T}^{(i)} \gets \displaystyle{\sum_{n=0}^{i-1}\sum_{j=0}^n {i-1\choose n }{n\choose j}{\bf T}^{(i-1-n)}{\bf R}^{(n-j+1)}{\bf T}^{(j)}}$
	\State $\displaystyle{f^{(i)}\gets \sum_{n=0}^{i-1}\sum_{j=0}^i {i-1\choose n}{n \choose j}(i-n) f^{(j)} f^{(i-j)}\delta^{(i-1-n)}}$
	\State $g^{(i)} \gets i\delta^{(i-1)}$
	\State $\delta^{(i)}\gets \frac{1}{K}\tr(\boldsymbol{\Phi}{\bf T}^{(i)})$
	\EndFor
	\caption{Iterative algorithm for computing ${\bf T}^{(q)}, \ q=1,\ldots,p$  }
\end{algorithmic}
\end{algorithm}

\section{Sketch of the proof of Theorem \ref{th:constraint}}
\label{app:scketch_c}
The goal of this section is to provide an outline of the proof for finding the deterministic equivalent of the quantity
$$
\left[\widetilde{\bf C}\right]_{\ell,m}=\frac{1}{K}\tr\left(\left(\frac{\wbH\wbHh}{K}\right)^\ell\wbH{\bf P}\wbHh\left(\frac{\wbH\wbHh}{K}\right)^m\right).
$$
A full proof proceeds in the following steps:
\begin{enumerate}
\item First compute the deterministic equivalent for 
$$
Y_{M}(t,u)=\frac{1}{K}\tr\left({\bf Q}(t)\wbH{\bf P}\wbHh {\bf Q}(u)\right)
$$
where $Q(t)=\left(\frac{t}{K}{\bf H}{\bf H}^{\mbox{\tiny H}}+{\bf I}\right)^{-1}$.
This can be achieved by using Lemma~\ref{lemma:new_deter}, where it is proved that
$$
Y_{M}(t,u)-\overline{\alpha}_M(t,u,{\bf I})\xrightarrow[M,K\to+\infty]{a.s}0
$$
and thus
$$
Y_{M}(t,u)-\tr({\bf P}) c(t,u)\xrightarrow[M,K\to+\infty]{\mathrm{a.s.}}0.
$$
\item Now, since
$$
\left[\widetilde{\bf C}\right]_{\ell,m}=\frac{(-1)^{\ell+m}Y_M^{(\ell,m)}}{\ell!m!}
$$
we can prove, using the same approach as in the proof of Theorem~\ref{corollary:derivative}, that
$$
Y_{M}(t,u)^{(\ell,m)}-\tr({\bf P}) c^{(\ell,m)}\xrightarrow[M,K\to+\infty]{a.s}0.
$$
\item Finally, one computes the derivative of $c(t,u)$ at $t=0$ and $u=0$, using the Leibniz rule, to arrive at the desired result.
\end{enumerate}

\section{Proof of Theorem~\ref{th:optimization}}
\label{app:optimization}
By using that $\frac{\tr \left({\bf P}\right){\bf w}^{\mbox{\tiny H}}\widetilde{\bf C}{\bf w}}{P}=1$ and dividing the objective function by the constant $\frac{K p_k}{\tr({\bf P})}$, the problem \eqref{eq:prob} can be rewritten as
\begin{align} \label{eq:problem}
	(P_1): \maximize{{\bf w}} & \quad \frac{{\bf w}^{\mbox{\tiny H}}\widetilde{\bf A}{\bf w}}{ {\bf w}^{\mbox{\tiny H}}\widetilde{\bf B}{\bf w}+
\frac{\sigma^2 }{P} {\bf w}^{\mbox{\tiny H}}\widetilde{\bf C}{\bf w}} \\
	 \mathrm{subject} \,\, \mathrm{to} & \quad \, {\bf w}^{\mbox{\tiny H}}\overline{\bf C}{\bf w}= \frac{P}{\tr \left({\bf P}\right)}. \notag
\end{align}
Making the change of variable ${\bf a}=\left(\widetilde{\bf B}+\frac{\sigma^2}{P}{\widetilde{\bf C}} \right)^{\frac{1}{2}} {\bf w}$, we transform $(P_1)$ into
\begin{align*}
	  &(P_2): \\
&\maximize{{\bf a}} \quad \frac{{\bf a}^{\mbox{\tiny H}}\left(\widetilde{\bf B}+\frac{\sigma^2}{P}\widetilde{\bf C}\right)^{\! -\frac{1}{2}}\widetilde{\bf A}\left(\widetilde{\bf B}+\frac{\sigma^2}{P}\widetilde{\bf C}\right)^{\! -\frac{1}{2}}{\bf a}}{{\bf a}^{\mbox{\tiny H}}{\bf a}} \\
&	 \mathrm{s.t.} \quad \,  {\bf a}^{\mbox{\tiny H}}\left(\widetilde{\bf B}+\frac{\sigma^2}{P}\widetilde{\bf C}\right)^{-\frac{1}{2}}\widetilde{\bf C}\left(\widetilde{\bf B}+\frac{\sigma^2}{P}\widetilde{\bf C}\right)^{\! -\frac{1}{2}}{\bf a}\!=\! \frac{P}{\tr \left({\bf P}\right)}.
\end{align*}
We notice that the objective function of $(P_2)$ is independent of the norm of ${\bf a}$. We can, therefore, select ${\bf a}$ to maximize the objective function and then adapt the norm to fit the constraint. If we discard the constraint, what remains is a classic Rayleigh quotient \cite{boyd}, which is maximized by the eigenvector ${\bf a}$ corresponding to the maximum eigenvalue of
$$
\left(\widetilde{\bf B}+\frac{\sigma^2}{P}\widetilde{\bf C}\right)^{-\frac{1}{2}}\widetilde{\bf A}\left(\widetilde{\bf B}+\frac{\sigma^2}{P}\widetilde{\bf C}\right)^{-\frac{1}{2}}.
$$
By transforming ${\bf a}$ back to the original variable ${\bf w}$ we obtain \eqref{eq:w-optimal}, where the scaling in \eqref{eq:alpha-optimal} corresponds to a scaling of ${\bf a}$ in order to satisfy the constraint.

\bibliographystyle{IEEEbib}
\bibliography{tutorial_RMT}

\begin{thebibliography}{10}

\bibitem{CISCO2013}
Cisco,
\newblock ``Cisco visual networking index: Global mobile data traffic forecast
  update, 2012-2017,''
\newblock {\em White Paper}, 2013.

\bibitem{Hoydis2011c}
J.~Hoydis, M.~Kobayashi, and M.~Debbah,
\newblock ``Green small-cell networks,''
\newblock vol. 6, no. 1, pp. 37--43, Mar. 2011.

\bibitem{Marzetta2010a}
T.L. Marzetta,
\newblock ``Noncooperative cellular wireless with unlimited numbers of base
  station antennas,''
\newblock vol. 9, no. 11, pp. 3590--3600, Nov. 2010.

\bibitem{Rusek2013a}
F.~Rusek, D.~Persson, B.K. Lau, E.G. Larsson, T.L. Marzetta, O.~Edfors, and
  F.~Tufvesson,
\newblock ``Scaling up {MIMO}: Opportunities and challenges with very large
  arrays,''
\newblock vol. 30, no. 1, pp. 40--60, Jan. 2013.

\bibitem{Hoydis2013a}
J.~Hoydis, S.~ten Brink, and M.~Debbah,
\newblock ``Massive {MIMO} in the {UL/DL} of cellular networks: How many
  antennas do we need?,''
\newblock vol. 31, no. 2, pp. 160--171, Feb. 2013.

\bibitem{Hosseini2013a}
K.~Hosseini, J.~Hoydis, S.~ten Brink, and M.~Debbah,
\newblock ``Massive {MIMO} and small cells: How to densify heterogeneous
  networks,''
\newblock in {\em Proc.~IEEE Int.~Conf.~Commun.~(ICC)}, 2013.

\bibitem{Bjornson2013e}
E.~Bj{\"{o}}rnson, M.~Kountouris, and M.~Debbah,
\newblock ``Massive {MIMO} and small cells: Improving energy efficiency by
  optimal soft-cell coordination,''
\newblock in {\em Proc.~Int.~Conf.~Telecommun.~(ICT)}, 2013.

\bibitem{Gao2011a}
X.~Gao, O.~Edfors, F.~Rusek, and F.~Tufvesson,
\newblock ``Linear pre-coding performance in measured very-large {MIMO}
  channels,''
\newblock in {\em Proc.~IEEE Veh. Tech. Conf. (VTC-Fall)}, 2011.

\bibitem{Hoydis2012a}
J.~Hoydis, C.~Hoek, T.~Wild, and S.~ten Brink,
\newblock ``Channel measurements for large antenna arrays,''
\newblock in {\em Int. Symp. Wireless Commun. Systems (ISWCS)}, 2012.

\bibitem{HAC06}
W.~Hachem, O.~Khorunzhy, P.~Loubaton, J.~Najim, and L.~A. Pastur,
\newblock ``{A new approach for capacity analysis of large dimensional
  multi-antenna channels},''
\newblock vol. 54, no. 9, pp. 3987--4004, Sept. 2008.

\bibitem{Nguyen2008a}
V.K. Nguyen and J.~Evans,
\newblock ``Multiuser transmit beamforming via regularized channel inversion: A
  large system analysis,''
\newblock in {\em Proc.~IEEE Global Commun.~Conf.~(GLOBECOM)}, 2008.

\bibitem{WAG10}
S.~Wagner, R.~Couillet, M.~Debbah, and D.~T.~M. Slock,
\newblock ``{Large System Analysis of Linear Precoding in MISO Broadcast
  Channels with Limited Feedback},''
\newblock vol. 58, no. 7, pp. 4509--4537, July 2012.

\bibitem{Muharar2011a}
R.~Muharar and J.~Evans,
\newblock ``Downlink beamforming with transmit-side channel correlation: A
  large system analysis,''
\newblock in {\em Proc.~IEEE Int.~Conf.~Commun.~(ICC)}, 2011.

\bibitem{COUbook}
R.~Couillet and M.~Debbah,
\newblock {\em {Random matrix methods for wireless communications}},
\newblock Cambridge University Press, New York, NY, USA, first edition, 2011.

\bibitem{Bjornson2012c}
E.~Bj{\"{o}}rnson, M.~Bengtsson, and B.~Ottersten,
\newblock ``Pareto characterization of the multicell {MIMO} performance region
  with simple receivers,''
\newblock vol. 60, no. 8, pp. 4464--4469, Aug. 2012.

\bibitem{PEE05}
C.~B. Peel, B.~M. Hochwald, and A.~L. Swindlehurst,
\newblock ``{A vector-perturbation technique for near-capacity multiantenna
  multiuser communication, Part I: Channel inversion and regularization},''
\newblock vol. 53, no. 1, pp. 195--202, Jan. 2005.

\bibitem{Lo1999a}
T.K.Y. Lo,
\newblock ``Maximum ratio transmission,''
\newblock vol. 47, no. 10, pp. 1458--1461, Oct. 1999.

\bibitem{Moshavi1996a}
S.~Moshavi, E.G. Kanterakis, and D.L. Schilling,
\newblock ``Multistage linear receivers for {DS-CDMA} systems,''
\newblock {\em Int. J. Wireless Information Networks}, vol. 3, no. 1, pp.
  1--17, Jan. 1996.

\bibitem{Honig2001a}
M.L. Honig and W.~Xiao,
\newblock ``Performance of reduced-rank linear interference suppression,''
\newblock vol. 47, no. 5, pp. 1928--1946, July 2001.

\bibitem{Sessler2005a}
G.~Sessler and F.~Jondral,
\newblock ``Low complexity polynomial expansion multiuser detector for {CDMA}
  systems,''
\newblock vol. 54, no. 4, pp. 1379--1391, July 2005.

\bibitem{hoydis}
{J. Hoydis and M. Debbah and M. Kobayashi},
\newblock ``{Asymptotic Moments for Interference Mitigation in Correlated
  Fading Channels},''
\newblock in {\em Proc. Int. Symp. Inf. Theory (ISIT)}, 2011.

\bibitem{Shariati2013a}
N.~Shariati, E.~Bj{\"o}rnson, M.~Bengtsson, and M.~Debbah,
\newblock ``Low-complexity channel estimation in large-scale {MIMO} using
  polynomial expansion,''
\newblock in {\em Proc.~IEEE Int.~Symp.~Personal, Indoor and Mobile Radio
  Commun.~(PIMRC)}, 2013.

\bibitem{Mueller2014b}
A.~M\"{u}ller, A.~Kammoun, E.~Bj\"{o}rnson, and M.~Debbah,
\newblock ``Efficient linear precoding for massive {MIMO} systems using
  truncated polynomial expansion,''
\newblock in {\em IEEE Sensor Array and Multichannel Signal Processing Workshop
  (SAM)}, 2014.

\bibitem{zarei}
S.~Zarei, W.~Gerstacker, R.~R. M\"uller, and R.~Schober,
\newblock ``{Low-Complexity Linear Precoding for Downlink Large-Scale MIMO
  Systems},''
\newblock in {\em Proc.~IEEE Int.~Symp.~Personal, Indoor and Mobile Radio
  Commun.~(PIMRC)}, 2013.

\bibitem{adhikary2013joint}
Ansuman Adhikary, Nam Junyoung, Jae-Young Ahn, and Giuseppe Caire,
\newblock ``{Joint Spatial Division and Multiplexing—The Large-Scale Array
  Regime},''
\newblock {\em IEEE transactions on information theory}, vol. 59, no. 10, pp.
  6441--6463, 2013.

\bibitem{Kammoun2014b}
A.~Kammoun, A.~M\"{u}ller, E.~Bj\"{o}rnson, and M.~Debbah,
\newblock ``Linear precoding based on polynomial expansion: Large-scale
  multi-cell mimo systems,''
\newblock Oct. 2014.

\bibitem{Choi2014a}
J.~Choi, D.J. Love, and P.~Bidigare,
\newblock ``Downlink training techniques for {FDD} massive {MIMO} systems:
  Open-loop and closed-loop training with memory,''
\newblock Sept. 2013,
\newblock Submitted, arXiv:1309.7712.

\bibitem{Wang2006a}
C.~Wang and R.D. Murch,
\newblock ``Adaptive downlink multi-user {MIMO} wireless systems for correlated
  channels with imperfect {CSI},''
\newblock vol. 5, no. 9, pp. 2435--2436, Sept. 2006.

\bibitem{Nosrat2011a}
B.~Nosrat-Makouei, J.G. Andrews, and R.W. Heath,
\newblock ``{MIMO} interference alignment over correlated channels with
  imperfect {CSI},''
\newblock vol. 59, no. 6, pp. 2783--2794, Jun. 2011.

\bibitem{Bjornson2013d}
E.~Bj{\"{o}}rnson and E.~Jorswieck,
\newblock ``Optimal resource allocation in coordinated multi-cell systems,''
\newblock {\em Foundations and Trends in Communications and Information
  Theory}, vol. 9, no. 2-3, pp. 113--381, 2013.

\bibitem{Joham2005a}
M.~Joham, W.~Utschick, and J.A. Nossek,
\newblock ``Linear transmit processing in {MIMO} communications systems,''
\newblock vol. 53, no. 8, pp. 2700--2712, Aug. 2005.

\bibitem{Sadek2007a}
M.~Sadek, A.~Tarighat, and A.H. Sayed,
\newblock ``A leakage-based precoding scheme for downlink multi-user {MIMO}
  channels,''
\newblock vol. 6, no. 5, pp. 1711--1721, May 2007.

\bibitem{Stridh2006a}
R.~Stridh, M.~Bengtsson, and B.~Ottersten,
\newblock ``System evaluation of optimal downlink beamforming with congestion
  control in wireless communication,''
\newblock vol. 5, no. 4, pp. 743--751, Apr. 2006.

\bibitem{Bjornson2010c}
E.~Bj{\"{o}}rnson, R.~Zakhour, D.~Gesbert, and B.~Ottersten,
\newblock ``Cooperative multicell precoding: Rate region characterization and
  distributed strategies with instantaneous and statistical {CSI},''
\newblock vol. 58, no. 8, pp. 4298--4310, Aug. 2010.

\bibitem{Shepard2012a}
C.~Shepard, H.~Yu, N.~Anand, L.E. Li, T.~Marzetta, R.~Yang, and L.~Zhong,
\newblock ``Argos: Practical many-antenna base stations,''
\newblock in {\em Proc.~ACM MobiCom}, 2012.

\bibitem{Boyd2008a}
S.~Boyd and L.~Vandenberghe,
\newblock ``Numerical~linear~algebra~background,''
\newblock http://www.ee.ucla.edu/ee236b/lectures/num-lin-alg.pdf.

\bibitem{Strassen1969a}
V.~Strassen,
\newblock ``Gaussian elimination is not optimal,''
\newblock {\em Numer. Math.}, vol. 13, pp. 354--356, 1969.

\bibitem{Williams2012a}
V.V. Williams,
\newblock ``Multiplying matrices faster than {Coppersmith-Winograd},''
\newblock in {\em Proc.~Symp.~Theory Comp.~(STOC)}, 2012, pp. 887--898.

\bibitem{Dahlman2011a}
E.~Dahlman, S.~Parkvall, and J.~Sk\"{o}ld,
\newblock {\em {4G}: {LTE/LTE}-Advanced for Mobile Broadband:
  {LTE/LTE}-Advanced for Mobile Broadband},
\newblock Academic Press, 2011.

\bibitem{Dick2007a}
C.~Dick, F.~Harris, M.~Pajic, and D.~Vuletic,
\newblock ``Implementing a real-time beamformer on an {FPGA} platform,''
\newblock in {\em Xcell J.}, 2007.

\bibitem{Bjornson2013f}
E.~Bj{\"{o}}rnson, E.~G. Larsson, and M.~Debbah,
\newblock ``Optimizing multi-cell massive {MIMO} for spectral efficiency: How
  many users should be scheduled?,''
\newblock in {\em GlobalSIP}, 2014,
\newblock Submitted.

\bibitem{Loyka2001a}
S.L. Loyka,
\newblock ``Channel capacity of {MIMO} architecture using the exponential
  correlation matrix,''
\newblock vol. 5, no. 9, pp. 369--371, 2001.

\bibitem{Golub1996a}
G.H. Golub and C.F.~Van Loan,
\newblock {\em Matrix Computations},
\newblock The Johns Hopkins University Press, 1996.

\bibitem{RUD86}
W.~Rudin,
\newblock {\em {Real and complex analysis}},
\newblock McGraw-Hill Series in Higher Mathematics, third edition, May 1986.

\bibitem{boyd}
S.~Boyd and L.~Vandenberghe,
\newblock {\em {Convex Optimization}},
\newblock Cambridge University Press, New York, 2004.

\end{thebibliography}

\end{document}